\documentclass[12pt]{article}
\usepackage[english]{babel}
\usepackage{amsmath,amsthm,amssymb,amsfonts,url}
\usepackage{graphicx,epstopdf,subfigure}
\usepackage{color}
\begin{document}

\title{Generalized Hamiltonian for a two-mode fermionic model and asymptotic equilibria}

\author{R.~D.~Salvo, M.~Gorgone and F.~Oliveri\\
\ \\
{\footnotesize Department of Mathematical and Computer Sciences,}\\
{\footnotesize Physical Sciences and Earth Sciences, University of Messina}\\
{\footnotesize Viale F. Stagno d'Alcontres 31, 98166 Messina, Italy}\\
{\footnotesize rdisalvo@unime.it; mgorgone@unime.it; foliveri@unime.it}
}

\date{Published in \textit{Physica A: Statistical Mechanics and its Applications} \textbf{540}, 123032 (2020).}

\maketitle

\begin{abstract}
In some recent papers, the so called  $(H,\rho)$-induced dynamics of a system $\mathcal{S}$ whose time evolution is deduced adopting an operatorial approach, has been introduced. According to the formal mathematical apparatus of quantum mechanics, $H$ denotes the Hamiltonian for $\mathcal{S}$, while $\rho$ is a certain rule applied periodically on $\mathcal{S}$. In this approach the rule acts at specific times $k\tau$, with $k$ integer and $\tau$ fixed, by modifying some of the parameters entering $H$ according to the state variation of the system. As a result, a dynamics admitting an asymptotic equilibrium state can be obtained. Here, we consider the limit for $\tau\rightarrow 0$, so that we introduce a generalized model leading to asymptotic equilibria. Moreover, in the case of a two-mode fermionic model, we
are able to derive a relation linking the parameters involved in the Hamiltonian 
to the asymptotic equilibrium states.
\end{abstract}


\section{Introduction}
\label{sec:intro}

Since 2006 \cite{bag2006}, raising and lowering operators typical of quantum mechanics \cite{Merzbacher,Messiah,roman} have been  successfully used for the mathematical modeling of several kinds of macroscopic systems.
Theoretical aspects related to methods and tools in the operatorial framework, together with many applications, are discussed in \cite{bagbook,bagbook_new}; moreover, many recent contributions in the area of quantum-like modeling outside physics, using the quantum formalism, and in particular the number representation, witness the flexibility of this general approach, which has proven effective in describing the dynamics in very different contexts, from social life and decision-making processes \cite{qdm1,qdm3,qdm4,qdm5,PK2016,dm1,dm2} to population migration  and crowd dynamics \cite{BO_migration,BGO_crowds,GAR14,BTBB}, but even ecological processes \cite{BO_ecomod,BCO_desert,DSO_RM2016,DSO_AAPP2016} or  political systems \cite{pol1,all1,all2,all3,BG17,DSO_turncoat2017,DSGO_turncoat2017,DSGO_opinion2017}. 

In general terms, the unknowns in an operatorial model are described by means of operators 
living in a Hilbert space that can be 
either finite or infinite dimensional (depending on the choice of using the fermionic rather 
than the bosonic representation). 
Once the observables of a macroscopic system $\mathcal{S}$, 
which are the self-adjoint operators relevant for the description of the system itself, have been identified,
the mean values of such operators computed on the eigenstate corresponding to an assigned initial 
condition give real valued functions that can be phenomenologically associated to some macroscopic quantities. This interpretation has revealed efficient in predicting the dynamics in many concrete situations \cite{bagbook,bagbook_new}.

The key aspect of the approach adopted here is that the time evolution of an observable 
$X$ of the macroscopic system $\mathcal{S}$ under consideration is given by 
$X(t)=\exp(\textrm{i}Ht)X\exp(-\textrm{i}Ht)$, where 
$H$ is some time independent self-adjoint Hamiltonian operator embedding the interactions occurring within the actors of the system. 
Not surprisingly, the description of the dynamics of operatorial models merely ruled by a time independent 
self-adjoint  Hamiltonian operator has some limitations; in fact, we obtain periodic or quasiperiodic evolutions. 
Then, if the system $\mathcal{S}$ we wish to model has some asymptotic \emph{final state}, 
it is clear that such a description does not work, and the modeling framework needs to be enriched, if 
not completely changed. This is not new: in quantum optics, or for two or three level atoms, if we need 
to describe a transition from one level to another, in some cases an effective finite dimensional 
non-Hermitian Hamiltonian has to be introduced \cite{Radmore}. In this way, decays are well described phenomenologically. 
Another way consists in considering the atoms interacting with some infinite reservoir, but  in such a 
case the full system ($\mathcal{S}_{full}$, \emph{i.e.}, $\mathcal{S}$ plus a suitable reservoir) certainly it is not finite dimensional. 

We should also mention that in the literature other ways are proposed to get some asymptotic behavior for a given system. The approach based on master equations, where the dynamics is described by some finite matrix whose entries are properly chosen, is one of this \cite{PK2016,ABKOTY2003,PKEH2016,BBBP2017}. This relies on the possibility of constructing a particularly simple dynamical equation, usually involving only few parameters. The model obtained in this way is easily solvable, but on the other hand several details of the interactions between the various agents of the original open system are lost.

Another possibility of obtaining asymptotic equilibria in operatorial models is based on the use of 
non-Hermitian Hamiltonian operators; see \cite{BGcancer} for a recent application 
to a dynamical system of tumor cells proliferation.

In \cite{BDSGO_GoL}, where a quantum version of game of life is considered, an extended version of the Heisenberg dynamics has been proposed in order to take into account effects which may occur during the time evolution of the system, and which can not apparently be included in a purely Hamiltonian description. In particular, during the evolution of the system, driven by a time independent Hermitian Hamiltonian $H$, at fixed times some checks on the system are performed, and these periodical measures on the system are used to change the state of the system itself according to an explicit prescription.  
A slightly different viewpoint has been proposed in \cite{DSO_AAPP2016,DSO_turncoat2017,DSGO_turncoat2017,DSGO_opinion2017}, where the periodical checks on the state of the system are used to change the values of some of the parameters entering the Hamiltonian operator, without modifying the functional form of the Hamiltonian itself. This approach proved to be quite efficient in operatorial models of stressed bacterial populations \cite{DSO_AAPP2016}, as well as in models of political systems affected by turncoat-like behaviors of part of their members,  {\em i.e.} systems characterized by internal fluxes between different political parties \cite{DSO_turncoat2017,DSGO_turncoat2017,DSGO_opinion2017}. In some sense, such an approach allows us to describe a sort of discrete self-adaptation of the model depending on the evolution of the state of the system, without the need of opening the system itself to any external reservoir, or considering more complex time dependent formulations. 

In \cite{BDSGO-PhysicaA}, the authors analyzed in general terms the approach of the so-called 
$(H,\rho)$-induced dynamics, and also provided a comparison with the well established frameworks where the 
Hamiltonian operator is time dependent or where the system is open and interacting with a reservoir. 
The strategy of the 
$(H,\rho)$-induced dynamics may give interesting results if the rule $\rho$ is not introduced as a mere 
mathematical trick, but  is somehow physically justified. For instance,  in \cite{DSO_AAPP2016}, the 
rule accounts for the modifications in the metabolic activity of bacteria due to lack of nutrients and/or to 
the presence of waste material, whereas, in 
\cite{DSO_turncoat2017,DSGO_turncoat2017,DSGO_opinion2017}, the rule modifies the behavior of 
the members of a political party with regard to their tendency to shift allegiance from one loyalty or ideal 
to another one.

In this paper, we describe the possibility of having a dynamics approaching an equilibrium state with
a generalized approach modeling a system that \emph{continuously adapts itself} according to its evolution. 
The procedure is illustrated by analyzing a rather simple model involving two fermionic operators, 
whose dynamics is ruled by a time independent, self-adjoint qua\-dratic Hamiltonian, according to 
the Heisenberg view. In such a situation, without any further ingredient, the resulting dynamics necessarily is periodic. If we consider the strategy of the $(H,\rho)$-induced dynamics (according to which the 
rule at fixed instants changes some of the parameters entering the Hamiltonian on the basis of the variations of 
the state of the system), the effect is that the system approaches asymptotically an 
equilibrium state. 
Then, we introduce a natural generalization of the rule by considering a suitable limit; in 
such a way, the parameters entering the Hamiltonian may gain a sort of dependence 
on the observables of the system, so that the corresponding values continuously change according to the 
instantaneous evolution of the system. This leads to the introduction of what we call a 
\emph{generalized Hamiltonian} embedding in its formulation the initial state of the system. 
Remarkably, by considering different sets of initial parameters, we are able to find a mathematical relation 
linking the values of the parameters entering the Hamiltonian with the value of the 
asymptotic equilibrium state.

The plan of the paper is the following. 
In section~\ref{sect2}, we present the two-mode fermionic model, and briefly review the strategy  of the 
$(H,\rho)$-induced dynamics according to which the rule acts on the parameters of the Hamiltonian at fixed 
instants. In section~\ref{sect3},  we introduce the generalized Hamiltonian, which is time independent 
and Hermitian, and exploit numerically the new approach. 
Section~\ref{sect4} is devoted to the derivation of the relation between the values of the parameters 
involved in the Hamiltonian and the asymptotic equilibrium state. 
Finally, section~\ref{sect5} contains our conclusions and perspectives.

\section{Two-mode fermionic system and $(H,\rho)$-in\-duced dynamics}
\label{sect2}

\begin{figure}[htb]
\centering
\includegraphics[width=0.7\textwidth]{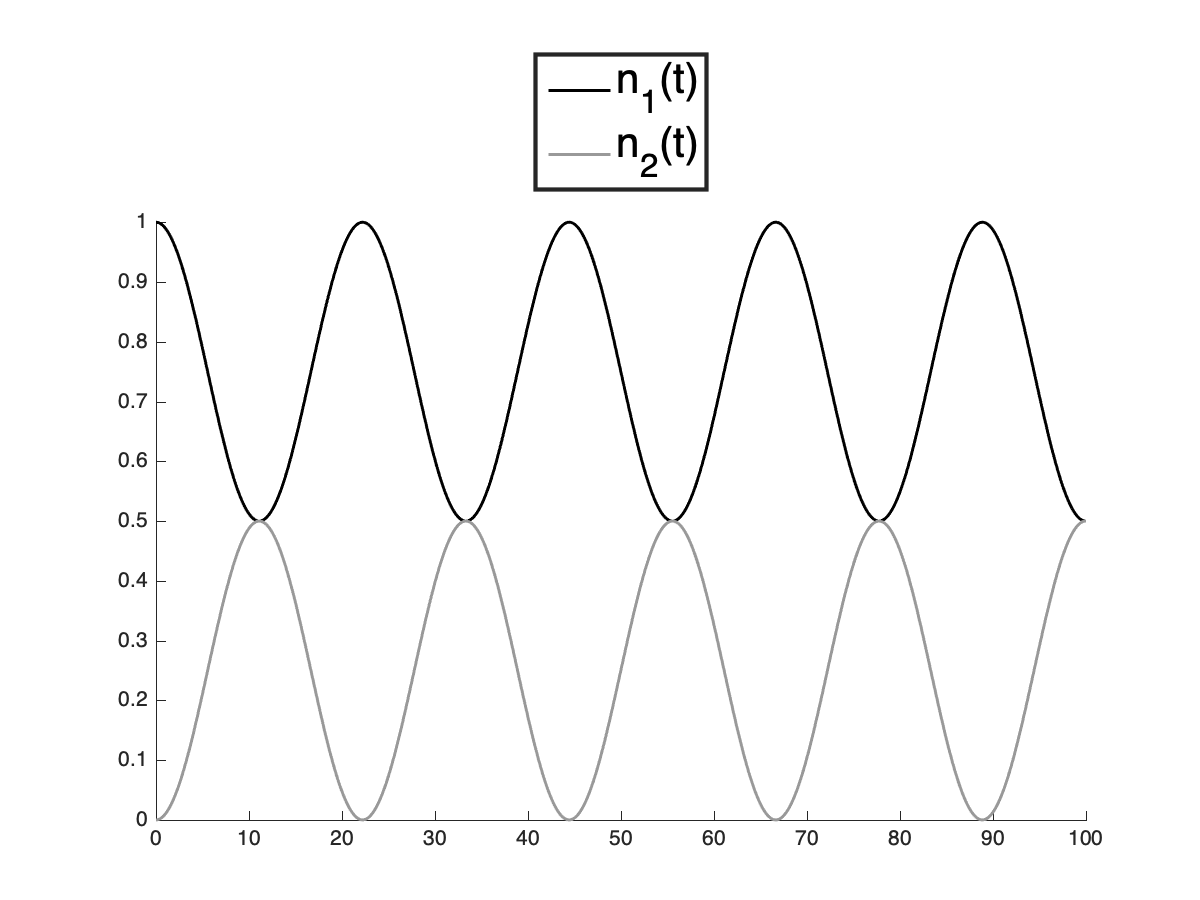}
\caption{\label{fig_norule}Plot of the solution \eqref{solmeanvalues} with $\omega_1=0.5$, 
$\omega_2=0.7$, $\lambda=0.1$, and
initial condition $\varphi_{1,0}$.}
\end{figure}

\begin{figure}
\centering
\subfigure[$\tau=1$]{\includegraphics[width=0.49\textwidth]{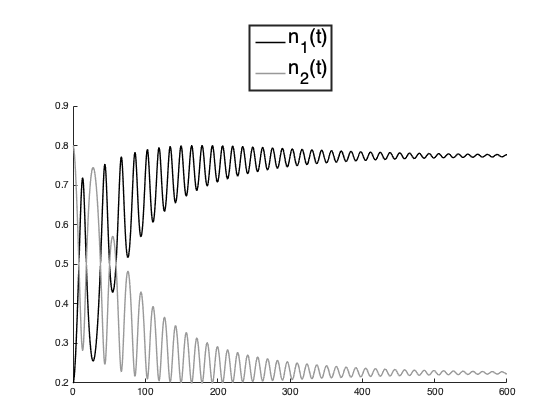}}
\subfigure[$\tau=2$]{\includegraphics[width=0.49\textwidth]{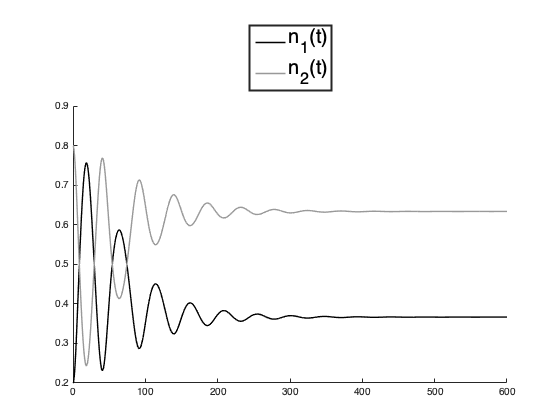}}
\subfigure[$\tau=3$]{\includegraphics[width=0.49\textwidth]{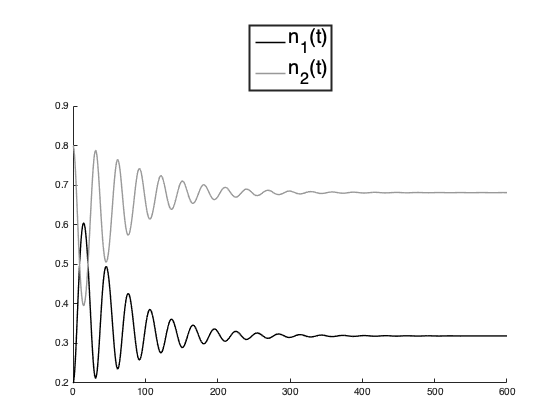}}
\subfigure[$\tau=4$]{\includegraphics[width=0.49\textwidth]{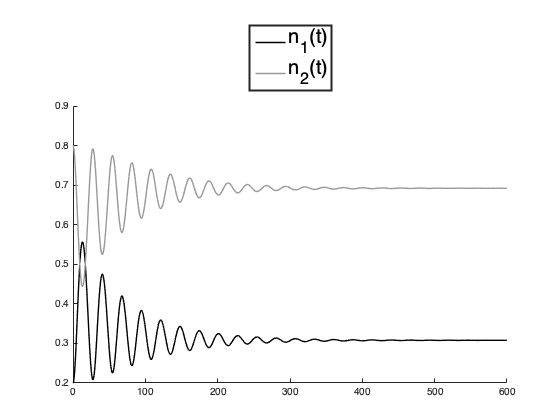}}
\subfigure[$\tau=6$]{\includegraphics[width=0.49\textwidth]{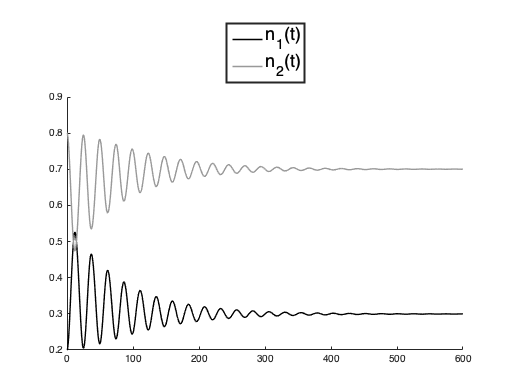}}
\subfigure[$\tau=10$]{\includegraphics[width=0.49\textwidth]{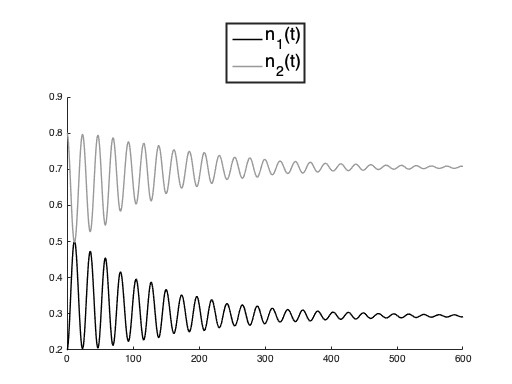}}
\caption{\label{fig:2mode_rule}Time evolution with rule \eqref{ruletau} 
for different choices of $\tau$; 
$\omega_1=0.5$, $\omega_2=0.7$, $\lambda=0.1$.}
\end{figure}

Consider a system $\mathcal{S}$ made by two interacting agents (or a system made by 
two compartments among which some fluxes occur).  Associate to each agent an annihilation ($a_j$), 
a creation ($a^\dagger_j$), and an occupation number ($\widehat n_j=a^\dagger_j a_j$) fermionic operator. 
Let us assume the dynamics to be governed by an observable operator $H$ (the Hamiltonian), that in 
standard quantum mechanics  corresponds to the total energy of the system. Its definition has to take 
account of the interactions among the parts of the system. 

The formalism of second quantization starts by postulating the Canonical Anticommutation Relations 
(CAR)
\begin{equation}
\label{eq:CAR}
\{a_j, a_k\}=0,\quad \{a_j^\dagger, a_k^\dagger\}=0,\quad
\{a_j, a_k^\dagger\}=\delta_{j,k}\mathbb{I},
\end{equation}
$j,k=1,2$, $\mathbb{I}$ being the identity operator, and $\{A,B\}:=AB+BA$ the anti-commutator 
between $A$ and $B$. 
Thus, since the Hilbert space $\mathcal{H}$ in which the fermionic system $\mathcal{S}$ 
lives is constructed as the linear span of the orthonormal set of vectors
\begin{equation}
\label{eq:fermion_vectors}
\varphi_{n_1,n_2}:=
(a_1^\dagger)^{n_1}(a_2^\dagger)^{n_2}\varphi_{0,0}, 
\end{equation} 
generated by acting on the \emph{vacuum} $\varphi_{0,0}$ with the operators 
$(a_\ell^\dagger)^{n_\ell}$, $n_\ell=0,1$ for $\ell=1,2$, it follows that the $\hbox{dim}(\mathcal{H})=4$.  
The vector $\varphi_{n_1,n_2}$ means that the first agent has a mean value equal to $n_1$, and the 
second one a mean value equal to $n_2$. We have 
\begin{equation} 
\widehat n_1\varphi_{n_1,n_2}=n_1\varphi_{n_1,n_2},\qquad \widehat
n_2\varphi_{n_1,n_2}=n_2\varphi_{n_1,n_2}. 
\label{MM22} 
\end{equation}

Let the dynamics of $\mathcal{S}$ be governed by the self-adjoint Hamiltonian
\begin{equation}
H=H_0+\lambda H_I,\quad H_0=\omega_1a_1^\dagger a_1+\omega_2a_2^\dagger a_2, 
\quad H_I=a_1^\dagger a_2+a_2^\dagger a_1,
\label{hamiltonian}
\end{equation}
where $\omega_j$ and $\lambda$ are real positive quantities. $H_0$ is the free part, and $\omega_1$, $\omega_2$ are parameters somehow related to the \emph{inertia} of the operators associated to the agents of $\mathcal{S}$: in fact, they measure the tendency of each degree of freedom to stay constant in time \cite{bagbook}. On the contrary, $H_I$ describes the interaction between the two agents by mimicking a sort of prey-predator mechanism.  
Of course, when $\lambda=0$, there is no contribution in $H$ due to the mutual interaction.

Adopting the Heisenberg view for the dynamics, the time evolutions of the annihilation operators 
$a_j(t)$ satisfy the differential equations
\begin{equation}
\label{dyneq2}
\begin{aligned}
&\dot a_1(t)=\textrm{i}[H,a_1]=\textrm{i} \left(-\omega_1 a_1(t) + \lambda a_2(t)\right) ,\\
&\dot a_2(t)=\textrm{i}[H,a_2]=\textrm{i} \left(\lambda a_1(t)-\omega_2 a_2(t)\right),
\end{aligned}
\end{equation}
that have to be solved with the initial conditions $a_1(0)={a_1}_0$ and $a_2(0)={a_2}_0$. 
The solution looks like
\begin{equation}
\begin{aligned}
&a_1(t)=\frac{1}{2\delta}\left({a_1}_0\left((\omega_1-\omega_2)\Phi_-(t)+\delta\Phi_+(t)\right)
+2\lambda {a_2}_0\Phi_-(t)\right),\\
&a_2(t)=\frac{1}{2\delta}\left({a_2}_0\left(-(\omega_1-\omega_2)\Phi_-(t)+\delta\Phi_+(t)\right)
+2\lambda {a_1}_0\Phi_-(t)\right),
\end{aligned}
\label{sol}
\end{equation}
where
\begin{equation}
\label{period}
\begin{aligned}
&\delta=\sqrt{(\omega_1-\omega_2)^2+4\lambda^2},\\
&\Phi_+(t)=2\exp\left(-\frac{\textrm{i}t(\omega_1+\omega_2)}{2}\right)\cos\left(\frac{\delta t}{2}\right),\\
&\Phi_-(t)=-2\textrm{i}\exp\left(-\frac{\textrm{i}t(\omega_1+\omega_2)}{2}\right)\sin\left(\frac{\delta t}{2}\right).
\end{aligned}
\end{equation}
Then, the functions 
$n_j(t):=\left<\varphi_{n_1,n_2},\widehat n_j(t)\varphi_{n_1,n_2}\right>$, giving the time evolution of the mean values of the number operators, are
\begin{equation}
\label{solmeanvalues}
\begin{aligned}
&n_1(t)=\frac{n_1(\omega_1-\omega_2)^2}{\delta^2}+ \frac{4\lambda^2}{\delta^2}
\left(n_1\cos^2\left(\frac{\delta t}{2}\right)+n_2\sin^2\left(\frac{\delta t}{2}\right)\right), \\
&n_2(t)=\frac{n_2(\omega_1-\omega_2)^2}{\delta^2}+ \frac{4\lambda^2}{\delta^2}
\left(n_2\cos^2\left(\frac{\delta t}{2}\right)+n_1\sin^2\left(\frac{\delta t}{2}\right)\right).
\end{aligned}
\end{equation}
These functions can be interpreted, in agreement with other applications considered in various concrete 
contexts, as the densities of two species, $\mathcal{S}_1$ and $\mathcal{S}_2$, interacting as in 
\eqref{hamiltonian} in a given (small) region \cite{BO_migration}. Since the interaction Hamiltonian $H_I$ in 
\eqref{hamiltonian} describes a sort of prey-predator mechanism, this reflects in the solution 
\eqref{solmeanvalues}, showing how the two densities, because of the interaction between  
$\mathcal{S}_1$ and $\mathcal{S}_2$, oscillate in the interval $[0,1]$
(see Figure~\ref{fig_norule}). 

If $\lambda=0$, the densities stay constant, $n_j(t)=n_j$, and nothing interesting happens in 
$\mathcal{S}$. We observe that the formulas in \eqref{solmeanvalues}
imply that $n_1(t)+n_2(t)=n_1+n_2$, independently of $t$ and the parameters 
$\omega_1$, $\omega_2$ and $\lambda$. We refer 
to \cite{BO_migration} for more details on this model, and for its role in modeling  migration
phenomena, which is achieved by considering a 2D version of the Hamiltonian in 
\eqref{hamiltonian} including an 
additional term in order to account for the diffusion of the two species in a lattice. 

In this paper, we exploit the possibility of getting some limiting values for $n_1(t)$ and $n_2(t)$ 
for large values of $t$, when $\lambda\neq 0$.

The first trivial remark is that the functions $n_1(t)$ and $n_2(t)$ in \eqref{solmeanvalues} do not admit 
any asymptotic limit, except when $n_1=n_2$ (or when $\lambda=0$, which is excluded here). In this 
case, clearly, $n_1(t)=n_2(t)=n_1=n_2$. On the other hand, if $n_1\neq n_2$, then both $n_1(t)$ and 
$n_2(t)$ always oscillate in time in opposition of phase. 

As shown in some recent papers, the description of the dynamics can be enriched by introducing a \emph{rule} able to
include in the model some effects that can not be embedded in the definition of $H$, unless we do not 
assume a time dependent Hamiltonian or include a reservoir. A simple rule is a law that modifies at fixed instants some of the values of the parameters involved in the Hamiltonian according to the current 
state of the system. In some sense, the underlying idea is that the model adjusts itself during the time evolution. 

Thus, let us consider again the model with two fermionic modes ruled by the Hamiltonian operator \eqref{hamiltonian}, 
and adopt the $(H,\rho)$-induced approach according to which the rule $\rho$ at fixed times $k\tau$  
($k=1,2,\ldots$), once chosen a positive value of $\tau$, modifies the parameters $\omega_1$ and $\omega_2$ in \eqref{hamiltonian}
according to the variations of the mean values $n_1$ and $n_2$ (computed on some assigned initial condition) in the time interval $[(k-1)\tau,k\tau]$ .

If we fix the initial values of the parameters, say  $\omega_1$, $\omega_2$, $\lambda$, the
initial conditions $\varphi_{n_1,n_2}$, and we assume the rule $\rho$ defined by
\begin{equation}
\label{ruletau}
\begin{aligned}
&\rho(\omega_1)=\omega_1(1+\delta_1(k)), \qquad &&\delta_1(k)=n_1(k\tau)-n_1((k-1)\tau),\\
&\rho(\omega_2)=\omega_2(1+\delta_2(k)), \qquad &&\delta_2(k)=n_2(k\tau)-n_2((k-1)\tau),
\end{aligned}
\end{equation}
we can see in  Figure~\ref{fig:2mode_rule} how the system tends to reach some asymptotic states in correspondence of different choices of $\tau$. 
We observe that the transient behavior of the time evolution changes with
$\tau$; in addition, also the value of the asymptotic equilibrium changes, at least for $\tau\le 3$. For larger values of $\tau$, the time needed to reach the equilibrium is increasing; 
this happens until $\tau$ approaches the value of 
the period of the solution that is obtained without the rule (for the chosen values of the parameters, according to formula \eqref{period}, this period is 22.2144), where no equilibrium is obtained.

The rule \eqref{ruletau} makes $\omega_1$ to increase (decrease, respectively) if 
$\delta_1(k)>0$ ($\delta_1(k)<0$, respectively). Notice that, since $n_1(t)+n_2(t)$ is a constant, 
$\delta_1(k)+\delta_2(k)\equiv0$, so that the variations of the inertia parameters 
$\omega_1$ and $\omega_2$ are opposite. Other choices are of course possible, depending
on the effects we want to include in the model. We observe that, for some sets of the initial parameters and different rules, it may happen that the system does not reach any equilibrium or 
takes a very long time to reach it.

The formalization of the $(H,\rho)$-induced approach can be summarized as follows. 
Let us start considering a self-adjoint quadratic Hamiltonian 
operator $H^{(1)}$; the corresponding evolution of a certain observable $X$ reads
\begin{equation}
X(t)=\exp(\textrm{i}H^{(1)}t)X\exp(-\textrm{i}H^{(1)}t),
\end{equation}
and its mean value computed on an eigenstate $\varphi_{n_1,n_2}$ is
\begin{equation}\label{add3}
x(t)=\langle\varphi_{n_1,n_2},\,
X(t)\varphi_{n_1,n_2}\rangle,
\end{equation}
for $t$ in a time interval of length $\tau>0$.
Then, let us modify some of the parameters involved in $H^{(1)}$, on the basis of the variations of the $x(t)$ computed according to \eqref{add3} after the time $\tau$ has elapsed. In this way, we get a new Hamiltonian operator $H^{(2)}$, having the same 
functional form as $H^{(1)}$, but (in general) with  different values of (some of) the involved parameters. 
Actually, we do not restart the evolution of the system from a new initial condition, but we simply continue to 
follow the evolution with the only difference that for $t\in]\tau,2\tau]$ the new Hamiltonian $H^{(2)}$ rules 
the process. And so on. The rule has to be thought of as a map acting on the space of the parameters 
involved in the Hamiltonian. Therefore, the 
global evolution is governed by a sequence of similar Hamiltonian operators, and the parameters 
entering the model can be considered stepwise (in time) constant.

To be more precise, let us consider a time interval $[0,T]$, and split it in $n=T/\tau$ 
subintervals of length $\tau$. Assume  $n$ to be integer. In the $k$-th subinterval $[(k-1)\tau,k\tau[$, consider a Hamiltonian $H^{(k)}$ ruling the dynamics. The global dynamics arises from the sequence of Hamiltonians
\begin{equation}
H^{(1)} \stackrel{\tau}{\longrightarrow} H^{(2)} \stackrel{\tau}{\longrightarrow} H^{(3)} 
\stackrel{\tau}{\longrightarrow} \ldots \stackrel{\tau}{\longrightarrow} H^{(n)},
\end{equation}
the complete evolution being obtained by glueing the local evolutions:
\begin{equation}
X(t)=\exp(\textrm{i}H^{(n)}t)\cdots\exp(\textrm{i}H^{(1)}t)X\exp(-\textrm{i}H^{(1)}t)\cdots\exp(-\textrm{i}H^{(n)}t).
\end{equation}

This rule-induced stepwise dynamics clearly may generate discontinuities in the first order 
derivatives of the operators, but prevents the occurrence of jumps in their evolutions and, consequently, 
in the mean values of the number operators.
By adopting this rule, we are implicitly considering the possibility of having a time dependent 
Hamiltonian (for a detailed comparison between the two approaches see \cite{BDSGO-PhysicaA}). A sort of time dependence is somehow hidden: in each subinterval 
$[(k-1)\tau,k\tau[$ the Hamiltonian does not depend on time, but at $k\tau$ some changes may occur, according to the system evolution. This means that, using this formulation, the evolution of the system is strongly influenced both by the rule and by the choice of the value of $\tau$. 
Here, our aim is to make the evolution of $\mathcal{S}$ independent of $\tau$: as a result, we can define a \emph{generalized Hamiltonian}, and still obtain a dynamics generally approaching an equilibrium state.

\section{A generalized Hamiltonian compatible with asymptotic equilibria}
\label{sect3}

\begin{figure}
\centering
\subfigure[$\boldsymbol{\alpha}=(0,-1)$]{\includegraphics[width=0.49\textwidth]{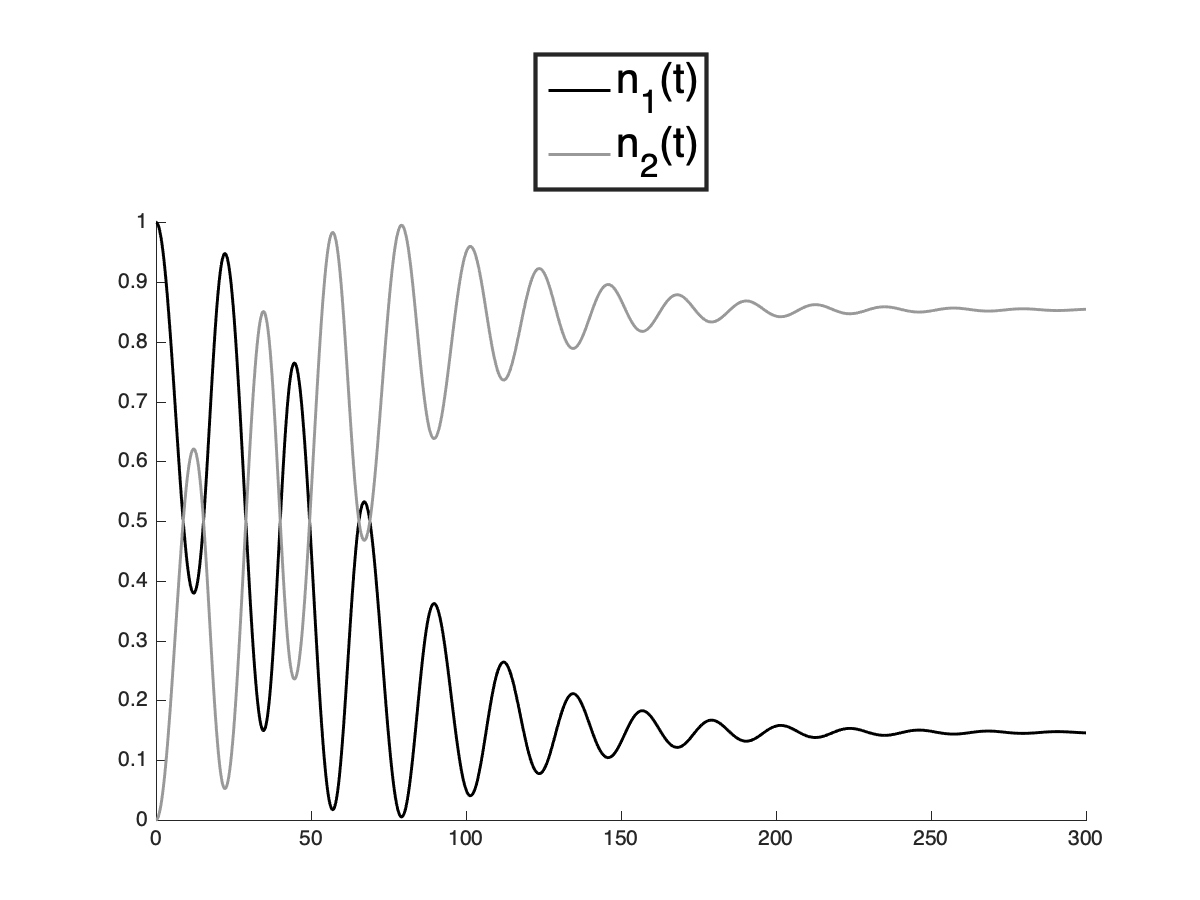}}
\subfigure[ $\boldsymbol{\alpha}=(-1,0)$]{\includegraphics[width=0.49\textwidth]{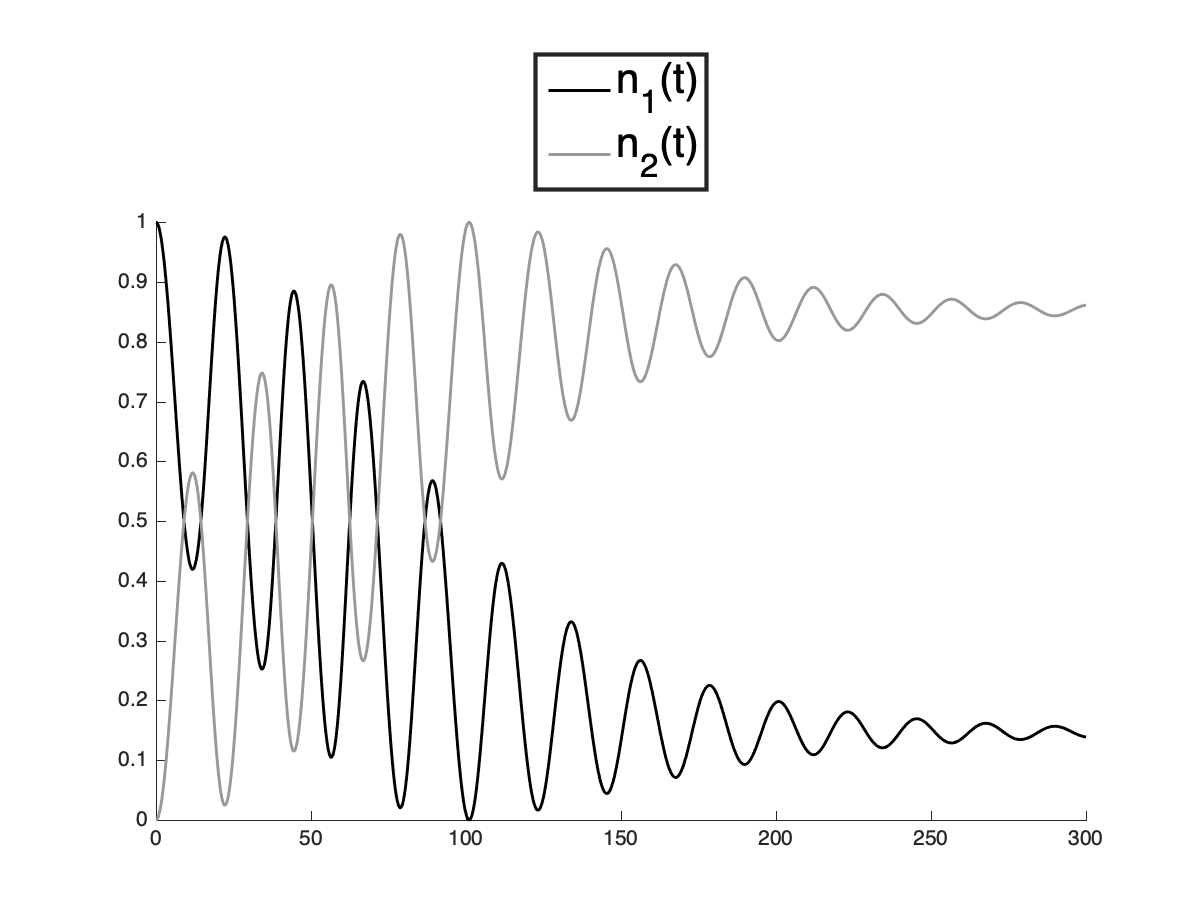}}\\
\subfigure[$\boldsymbol{\alpha}=(0,1)$]{\includegraphics[width=0.49\textwidth]{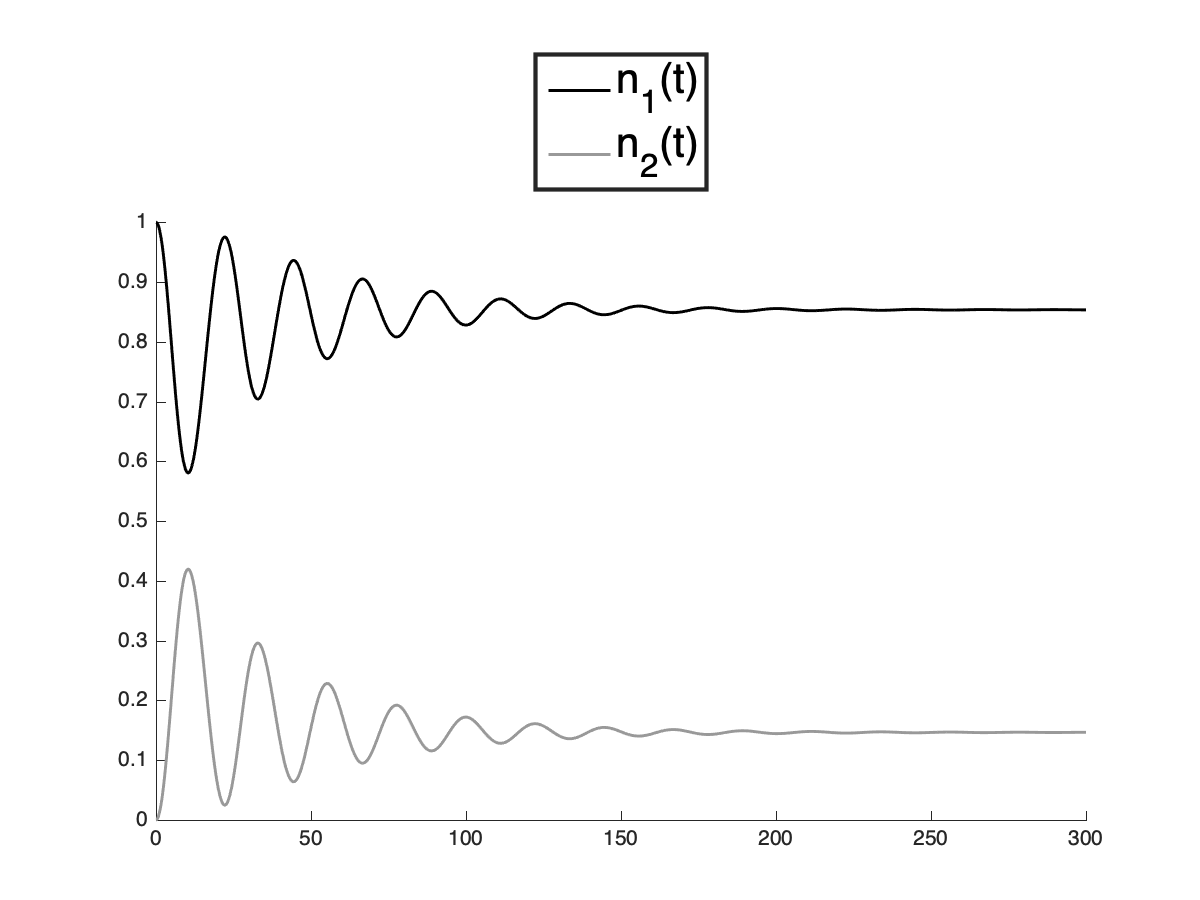}}
\subfigure[$\boldsymbol{\alpha}=(1,0)$]{\includegraphics[width=0.49\textwidth]{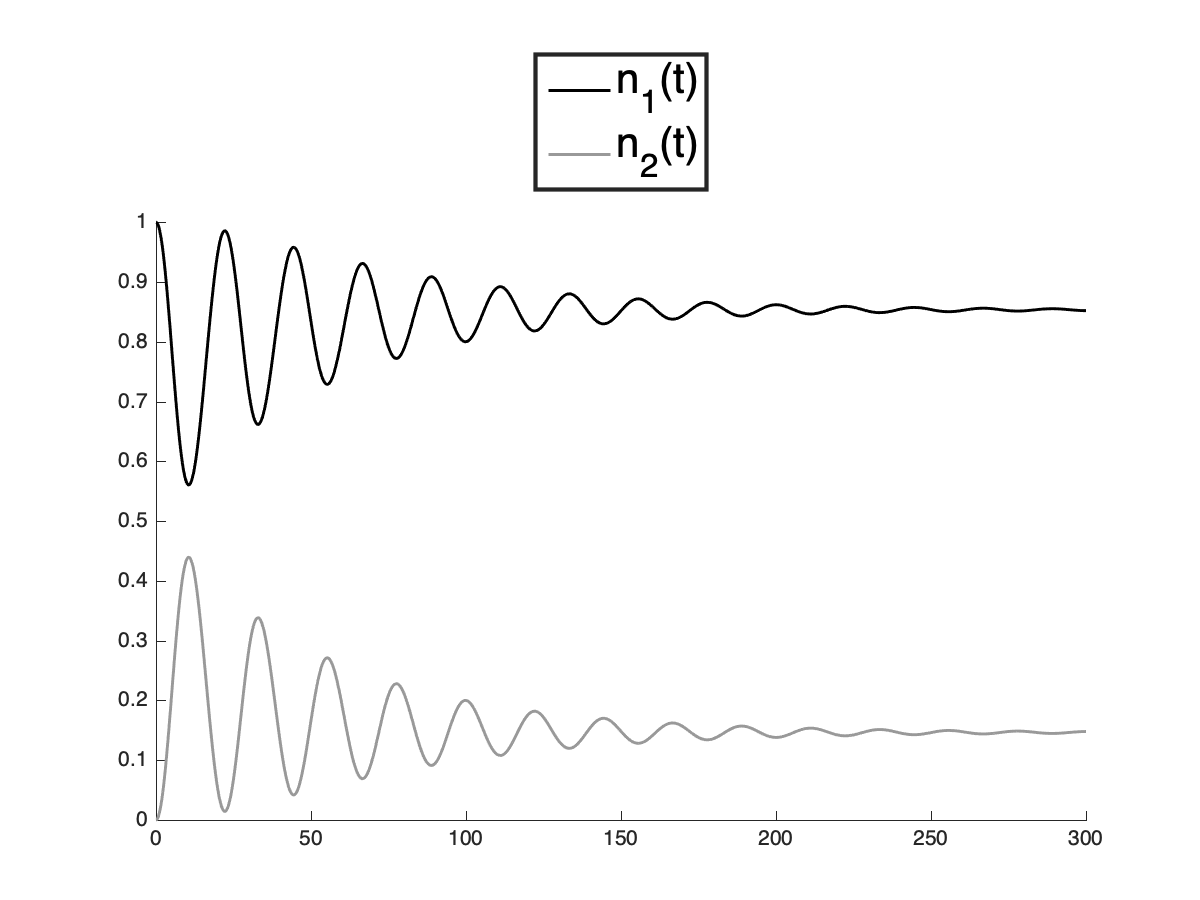}}\\
\subfigure[$\boldsymbol{\alpha}=(-1,-1)$]{\includegraphics[width=0.49\textwidth]{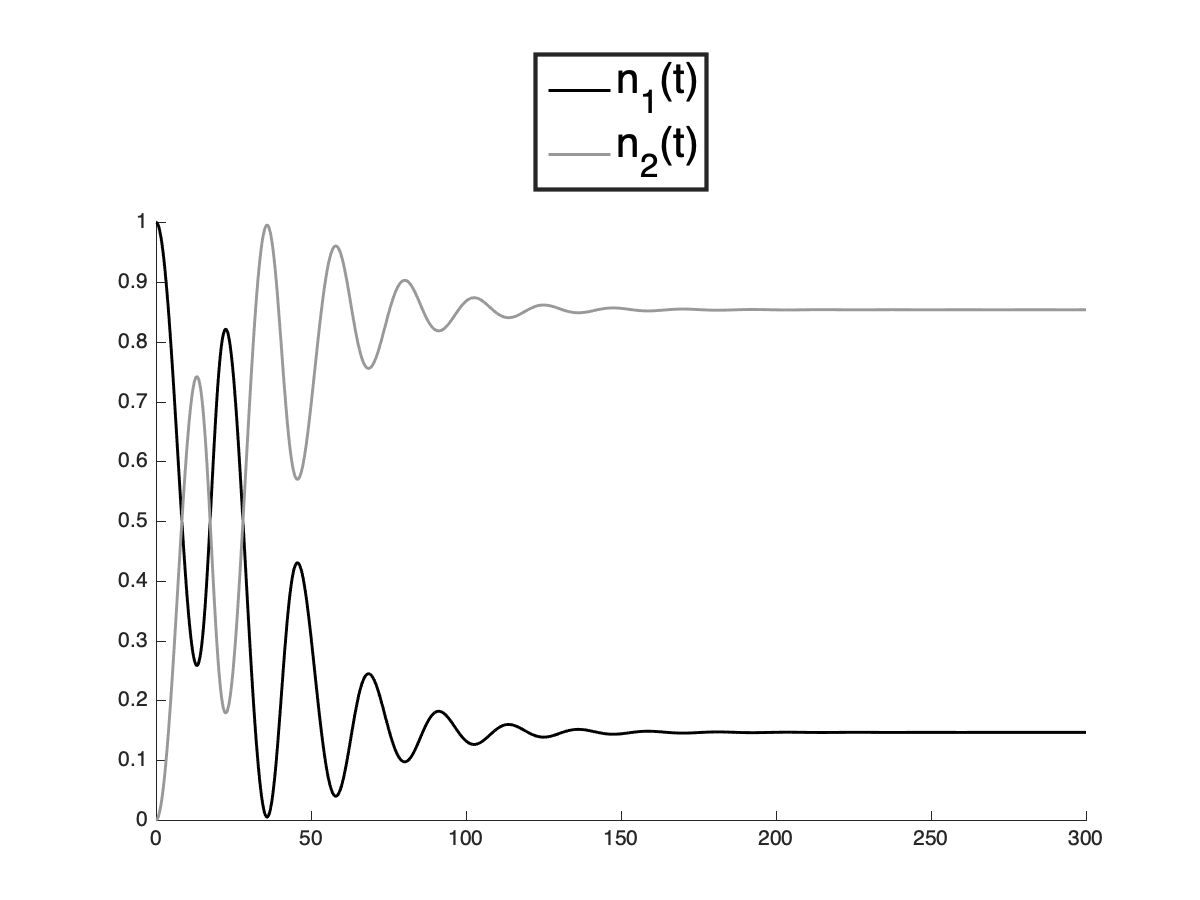}}
\subfigure[$\boldsymbol{\alpha}=(1,1)$]{\includegraphics[width=0.49\textwidth]{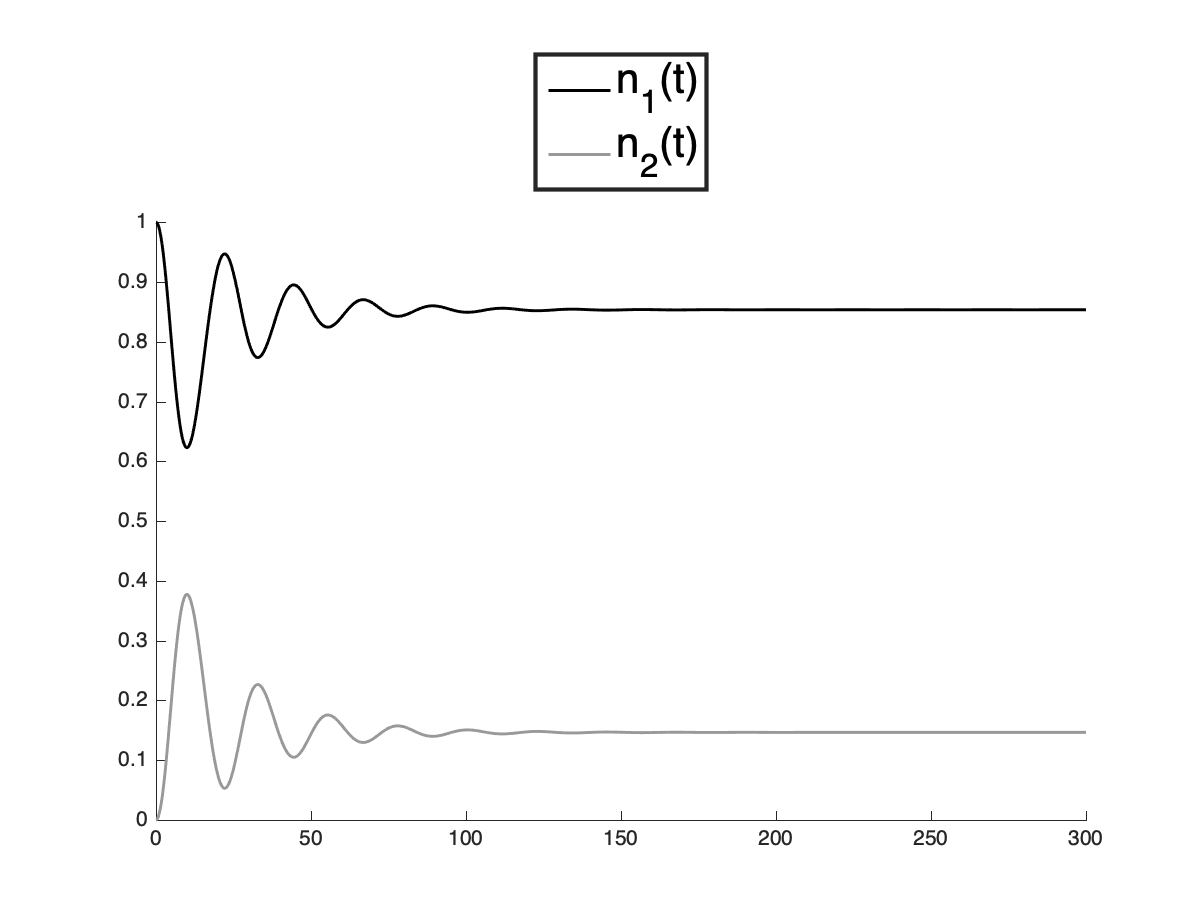}}\\
\caption{\label{fig_om1_lt_om2_lam_01}Plots of the solutions of equations \eqref{eq_rhocontinuous} with
$\lambda=0.1$ and the initial values of the inertia parameters ${\omega_1}_0=0.5$, ${\omega_2}_0=0.7$
according to various choices of $\alpha_1$ and $\alpha_2$.}
\end{figure}

\begin{figure}
\centering
\subfigure[$\boldsymbol{\alpha}=(-1,1)$]{\includegraphics[width=0.49\textwidth]{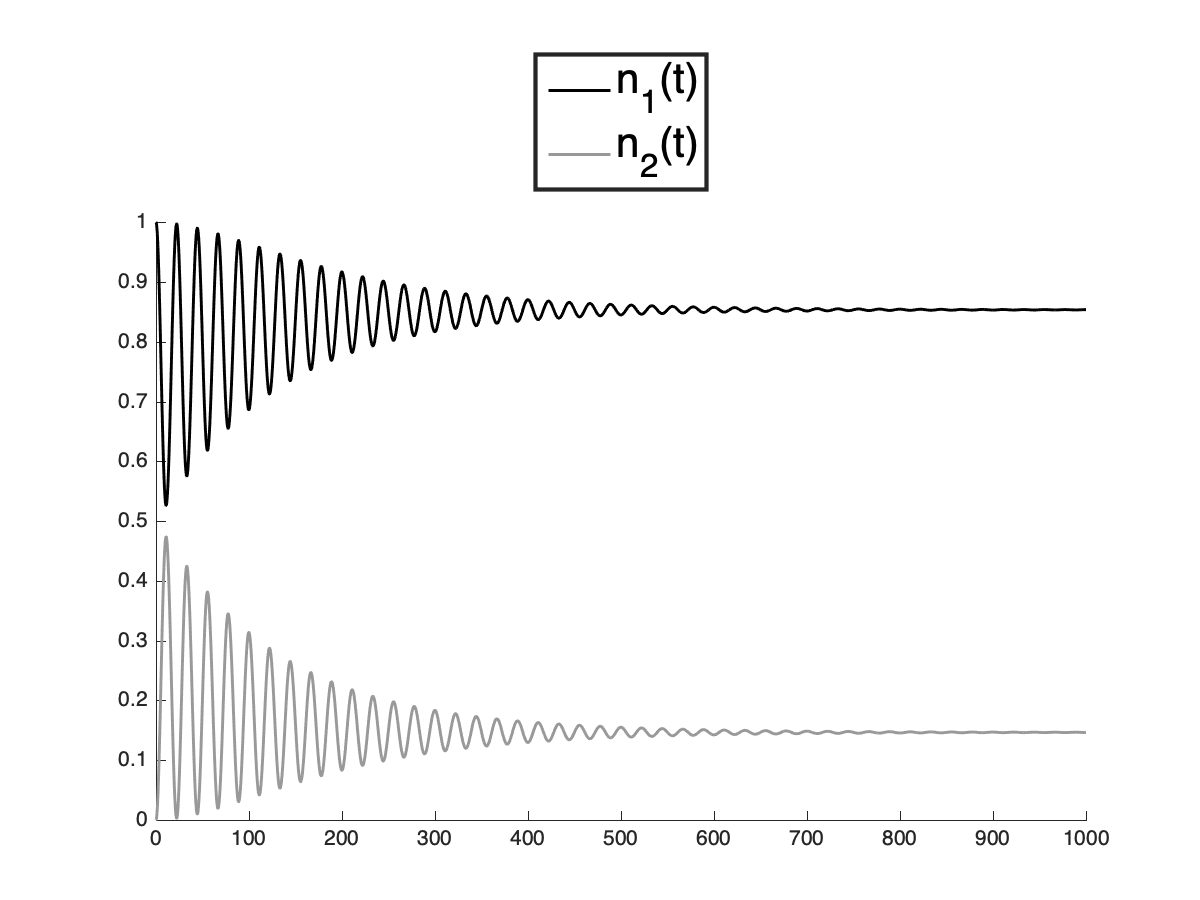}}
\subfigure[$\boldsymbol{\alpha}=(1,-1)$]{\includegraphics[width=0.49\textwidth]{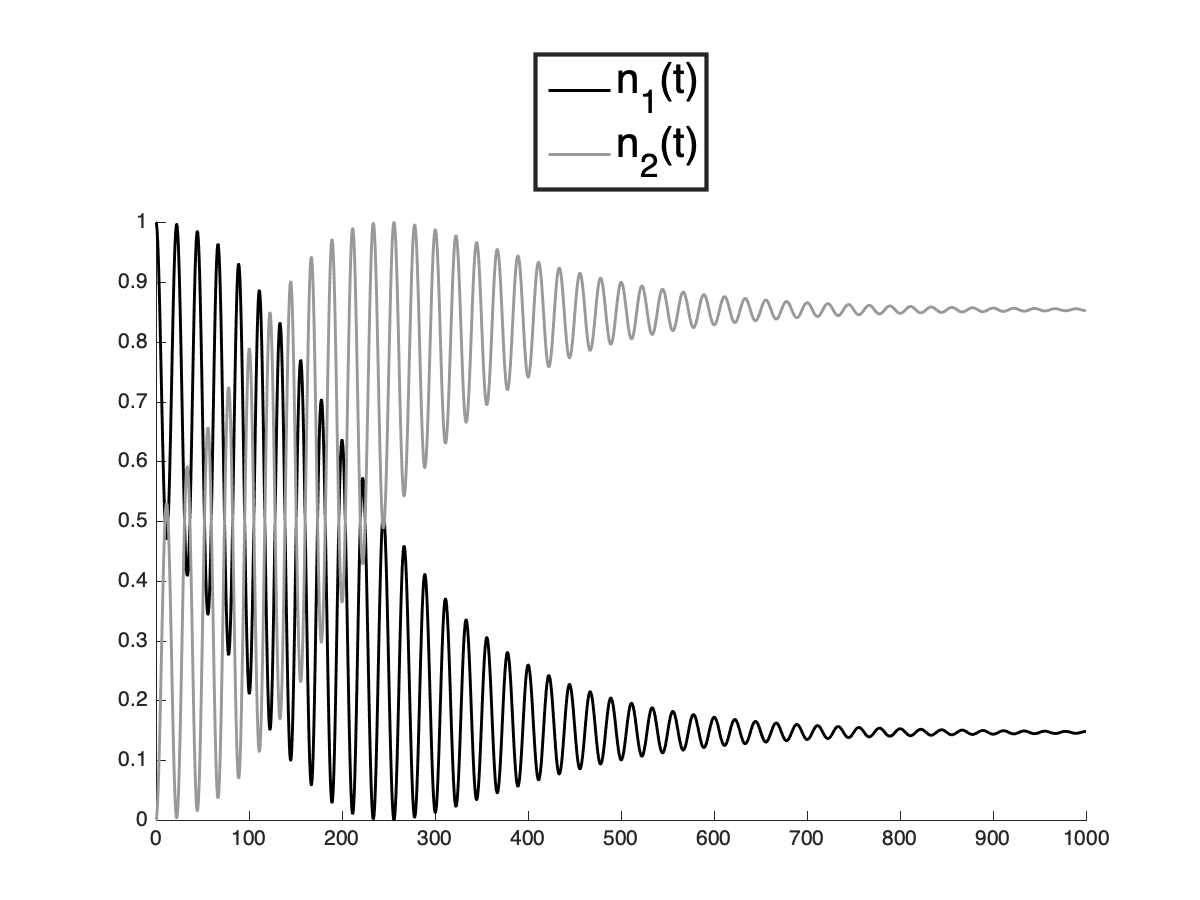}}
\caption{\label{fig_om1_lt_om2_lam_01_bis}Plots of the solutions of equations \eqref{eq_rhocontinuous} with
$\lambda=0.1$ and the initial values of the inertia parameters ${\omega_1}_0=0.5$, ${\omega_2}_0=0.7$
when $\alpha_1\alpha_2=-1$.}
\end{figure}

\begin{figure}
\centering
\subfigure[$\boldsymbol{\alpha}=(0,-1)$]{\includegraphics[width=0.49\textwidth]{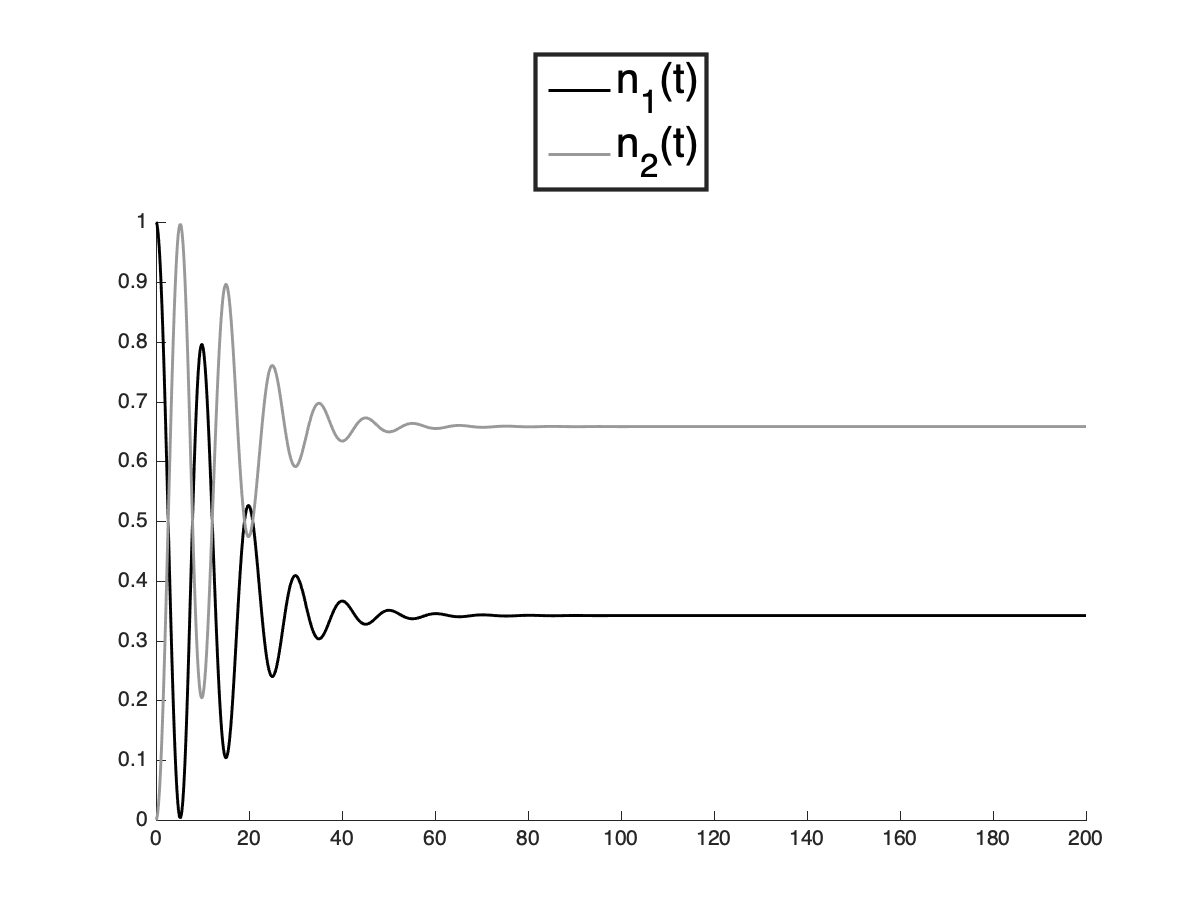}}
\subfigure[ $\boldsymbol{\alpha}=(-1,0)$]{\includegraphics[width=0.49\textwidth]{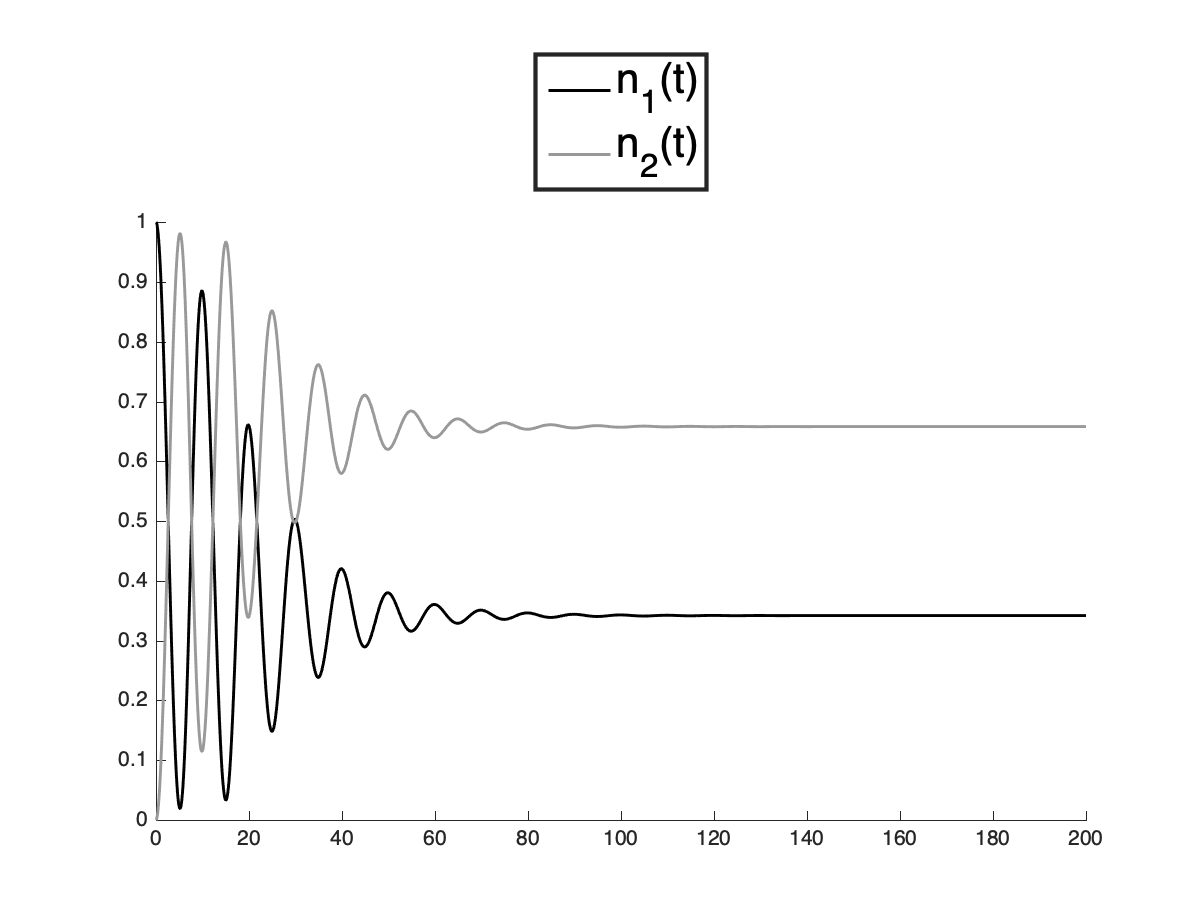}}\\
\subfigure[$\boldsymbol{\alpha}=(0,1)$]{\includegraphics[width=0.49\textwidth]{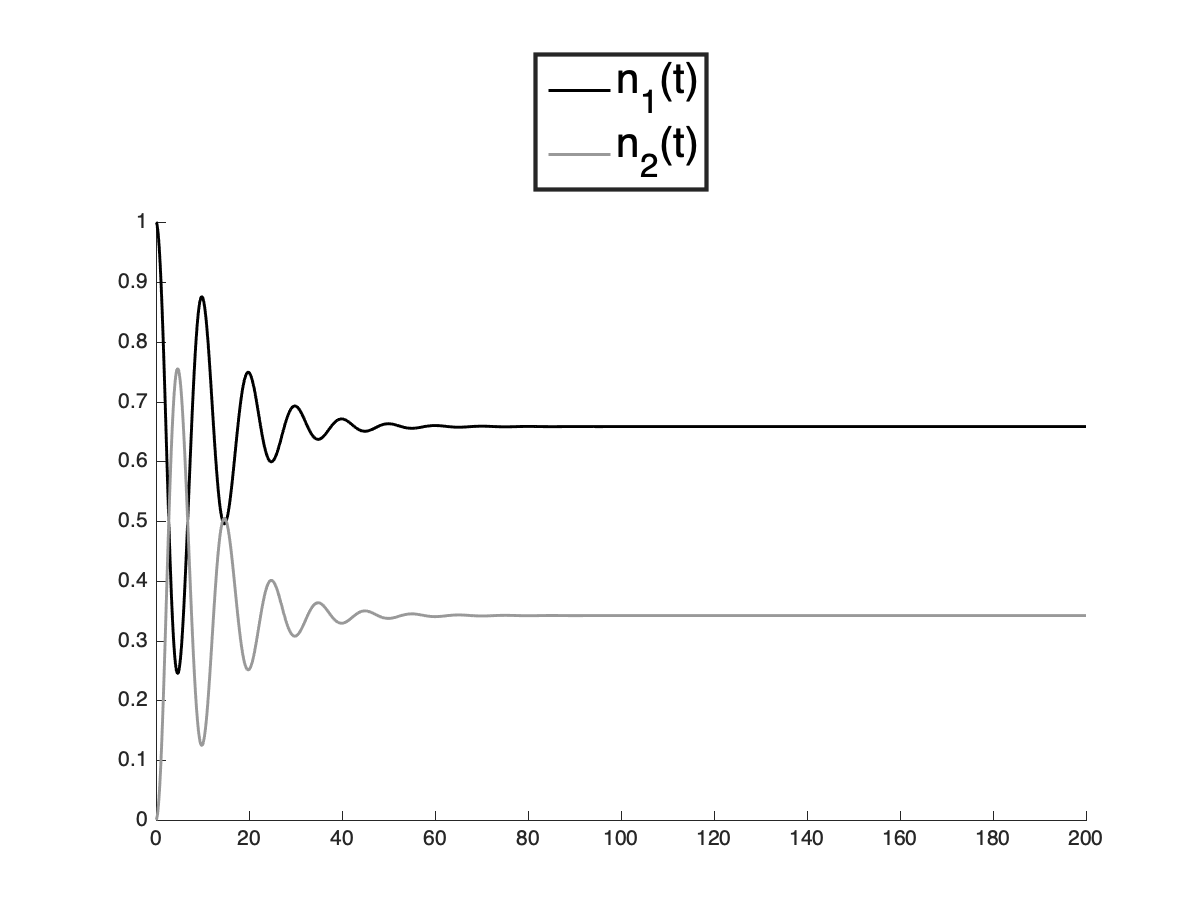}}
\subfigure[$\boldsymbol{\alpha}=(1,0)$]{\includegraphics[width=0.49\textwidth]{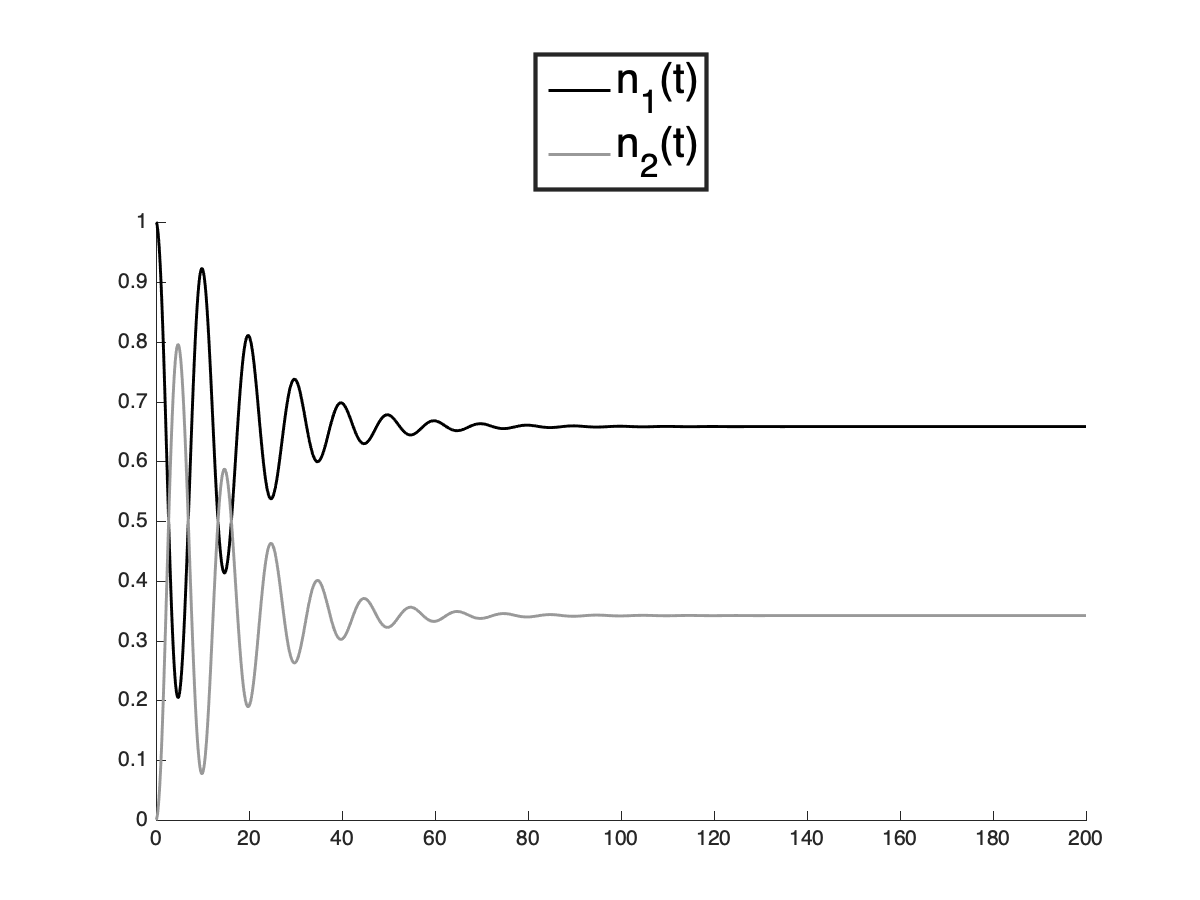}}\\
\subfigure[$\boldsymbol{\alpha}=(-1,-1)$]{\includegraphics[width=0.49\textwidth]{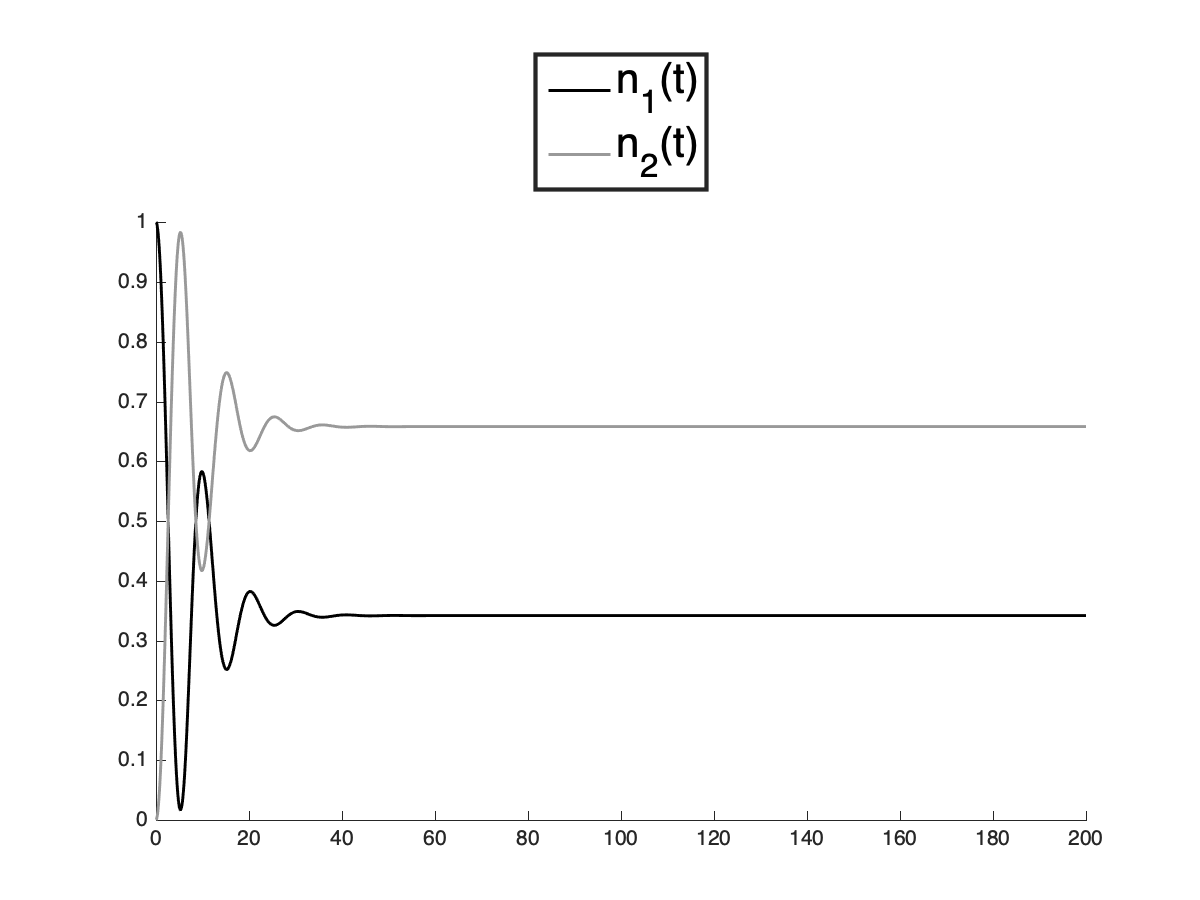}}
\subfigure[$\boldsymbol{\alpha}=(1,1)$]{\includegraphics[width=0.49\textwidth]{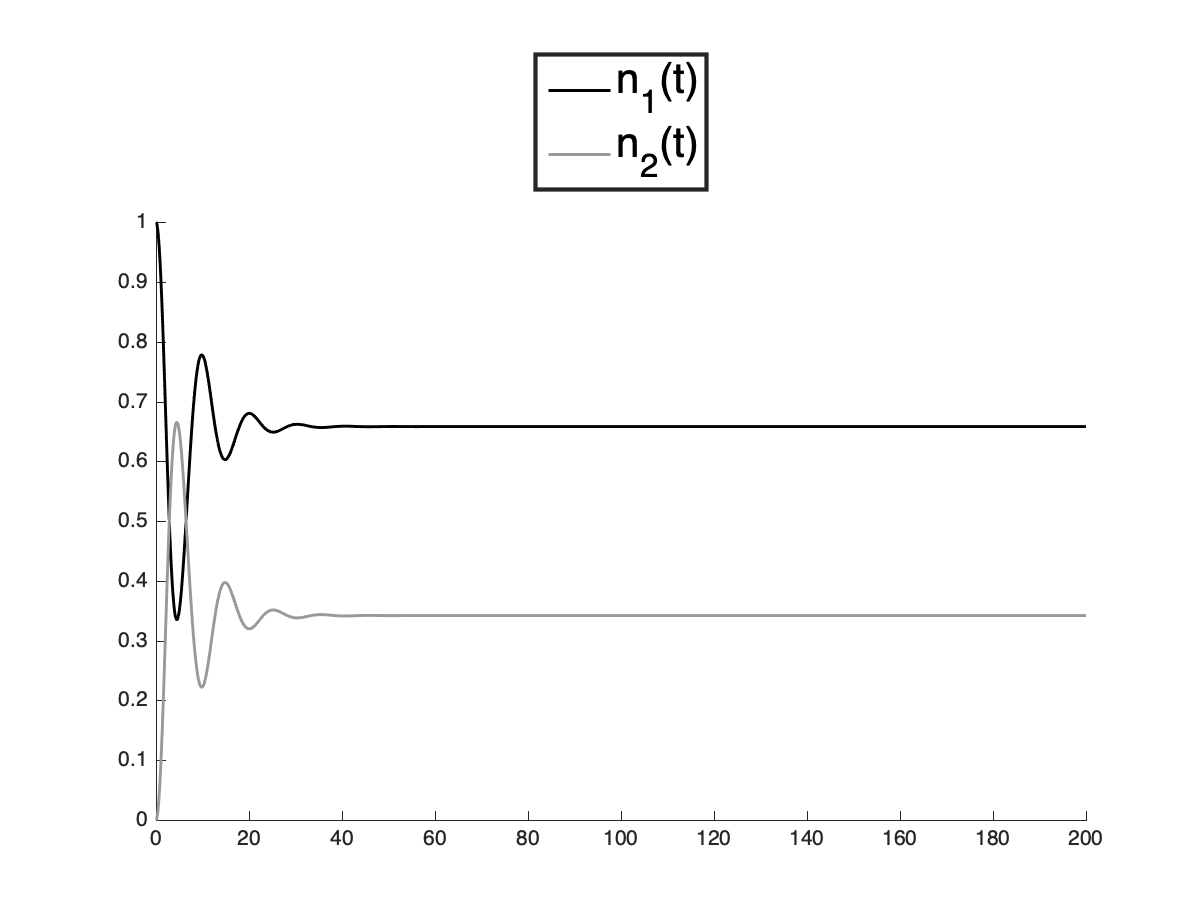}}\\
\caption{\label{fig_om1_lt_om2_lam_03}Plots of the solutions of equations \eqref{eq_rhocontinuous} with
$\lambda=0.3$ and the initial values of the inertia parameters ${\omega_1}_0=0.5$, ${\omega_2}_0=0.7$
according to various choices of $\alpha_1$ and $\alpha_2$.}
\end{figure}

\begin{figure}
\centering
\subfigure[$\boldsymbol{\alpha}=(-1,1)$]{\includegraphics[width=0.49\textwidth]{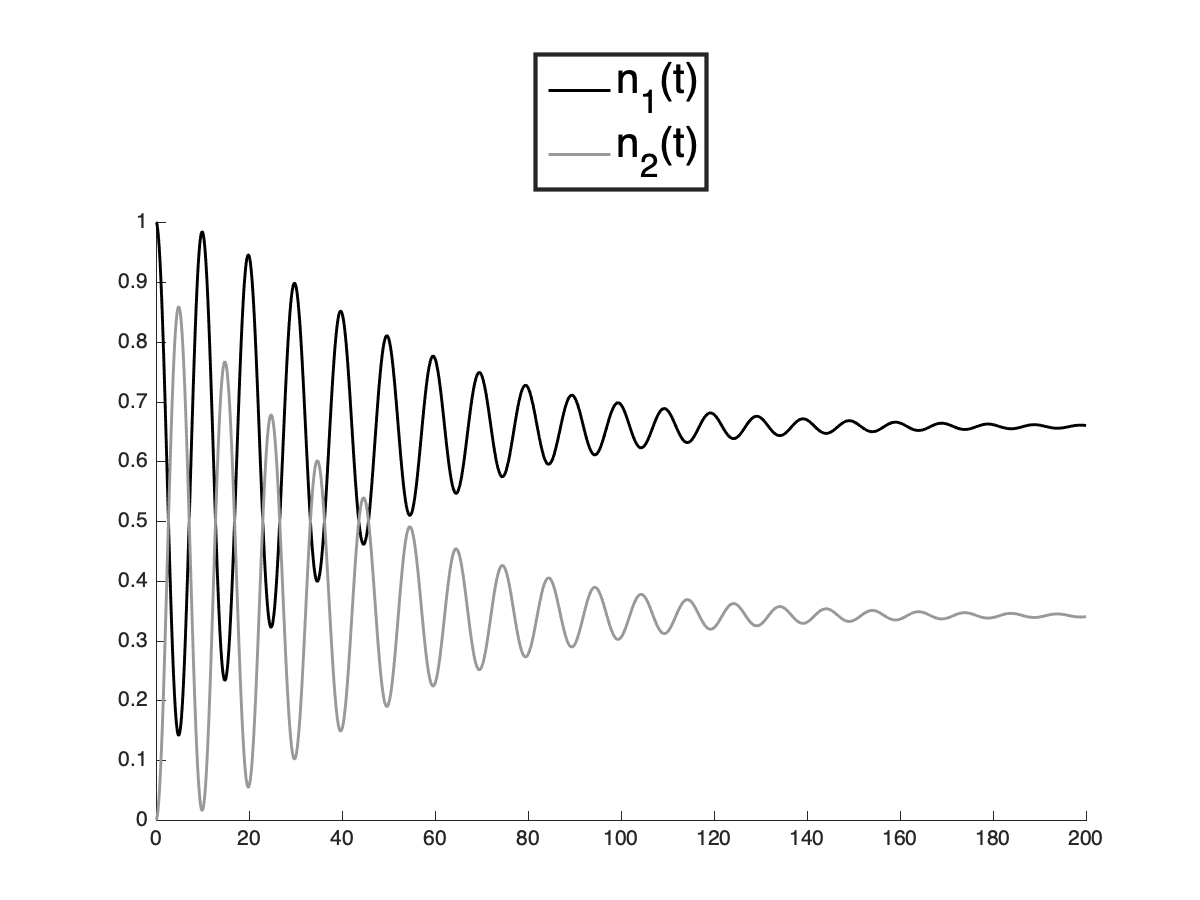}}
\subfigure[$\boldsymbol{\alpha}=(1,-1)$]{\includegraphics[width=0.49\textwidth]{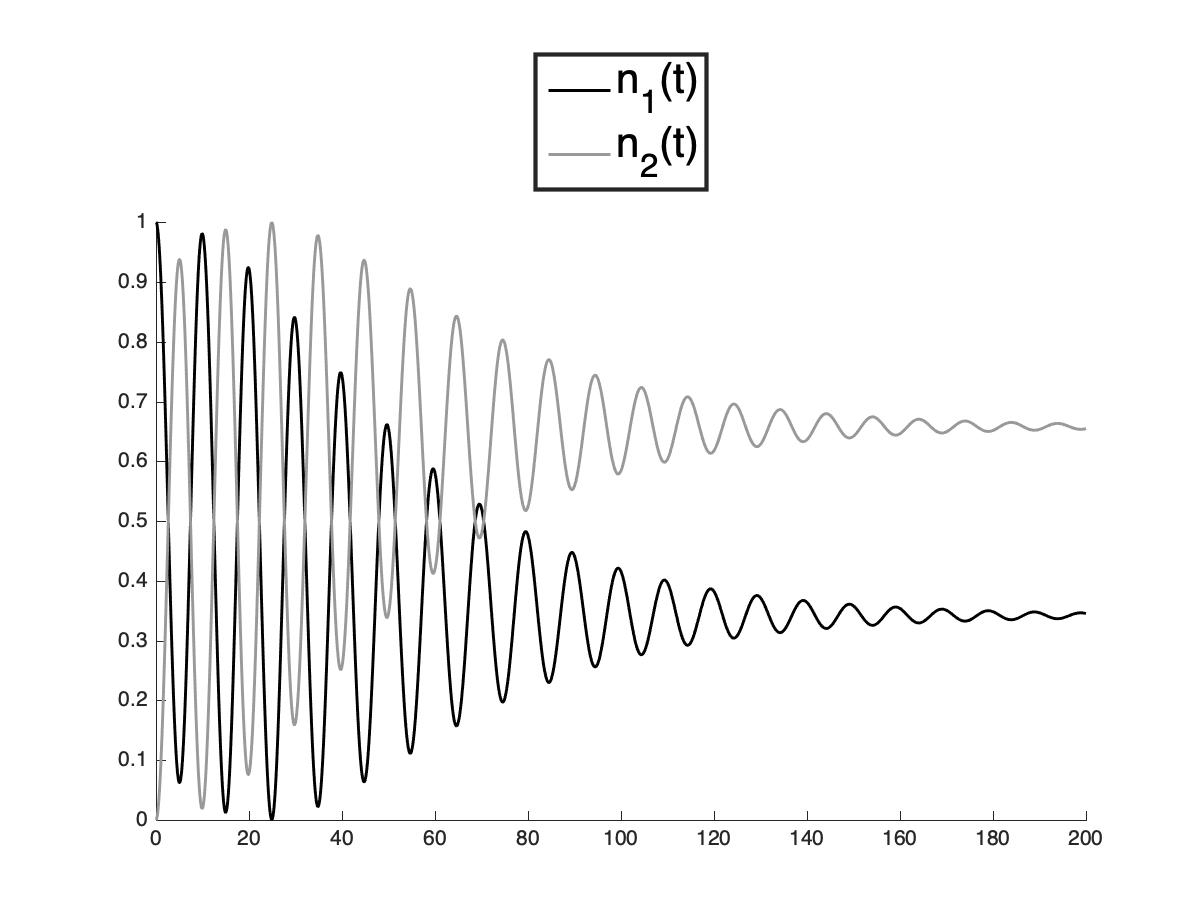}}
\caption{\label{fig_om1_lt_om2_lam_03_bis}Plots of the solutions of equations \eqref{eq_rhocontinuous} with
$\lambda=0.3$ and the initial values of the inertia parameters ${\omega_1}_0=0.5$, ${\omega_2}_0=0.7$
when $\alpha_1\alpha_2=-1$.}
\end{figure}

\begin{figure}
\centering
\subfigure[$\boldsymbol{\alpha}=(0,-1)$,$\boldsymbol{\alpha}=(-1,0)$]{\includegraphics[width=0.49\textwidth]{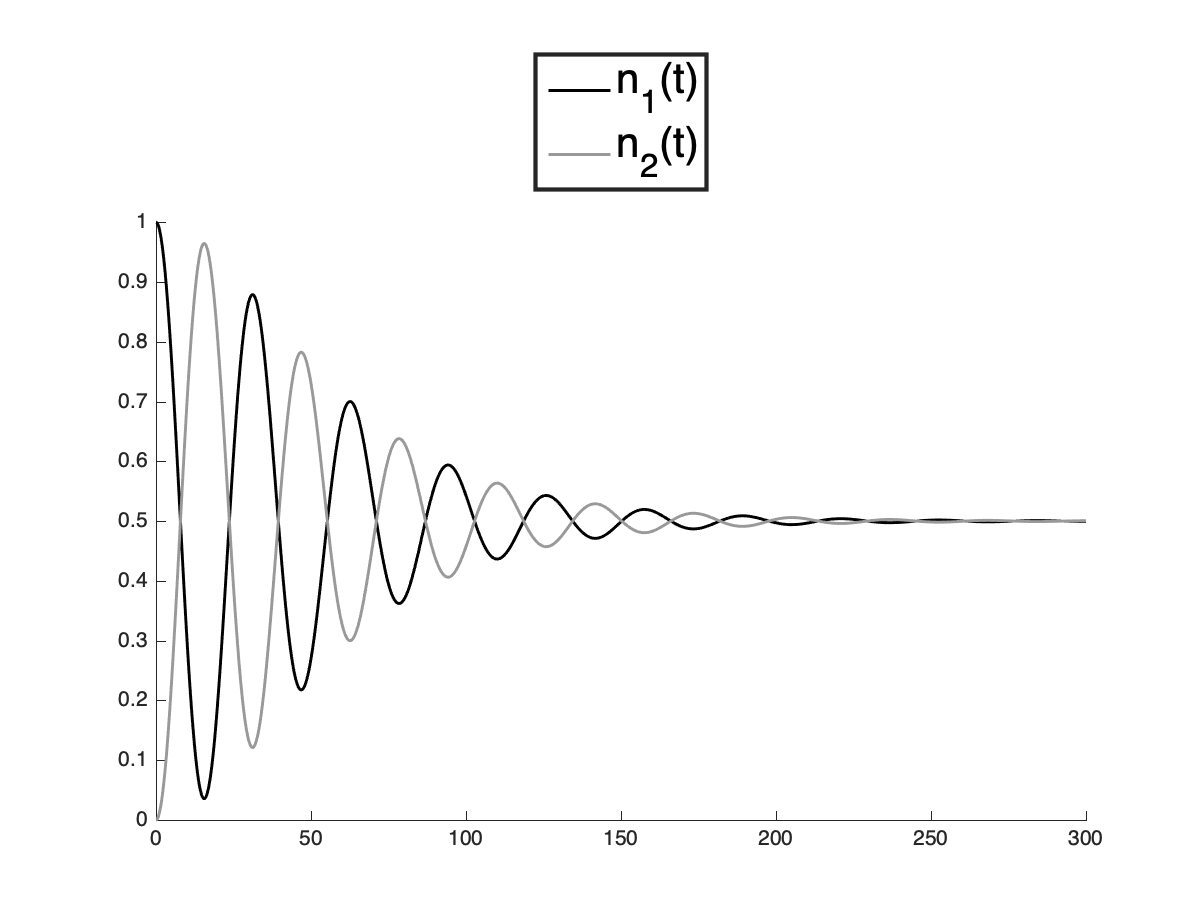}}
\subfigure[$\boldsymbol{\alpha}=(1,0)$, $\boldsymbol{\alpha}=(1,0)$]{\includegraphics[width=0.49\textwidth]{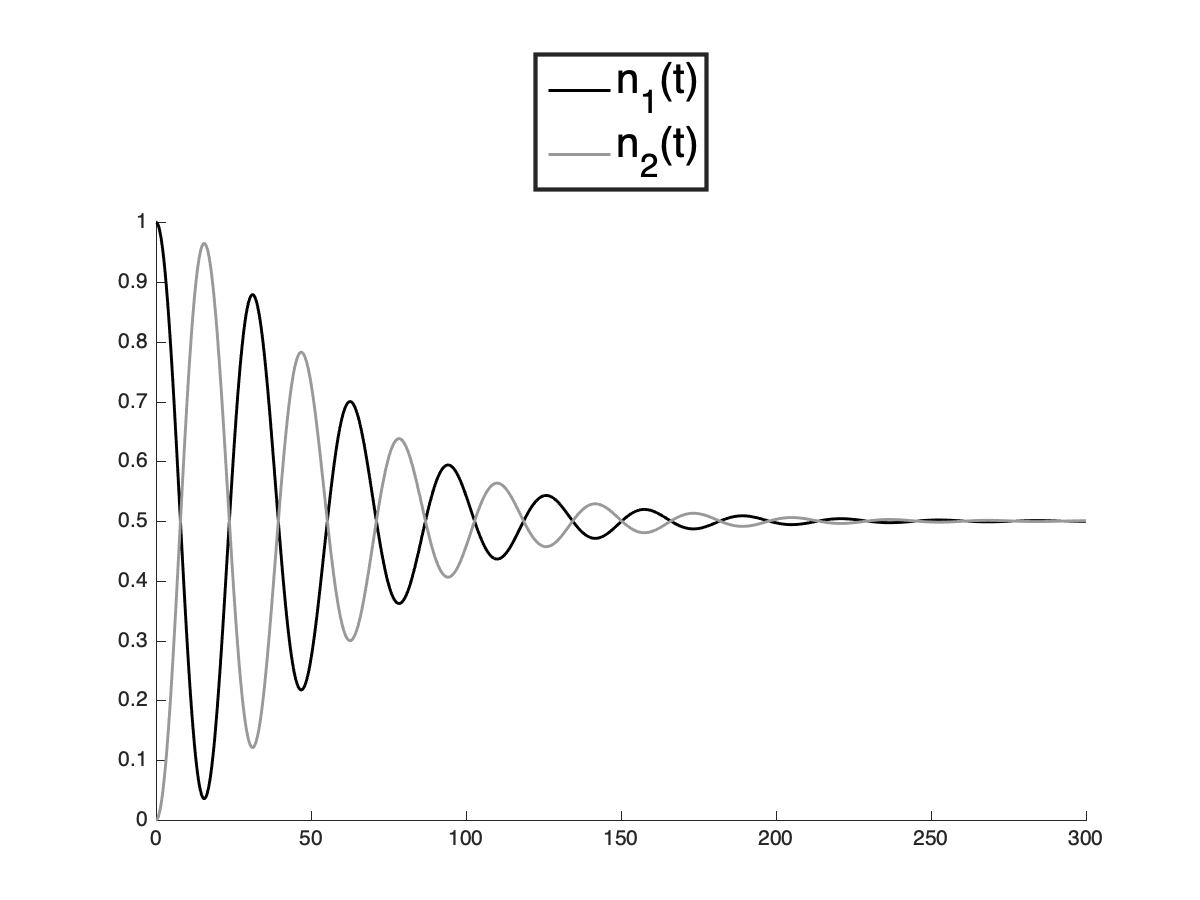}}\\
\subfigure[$\boldsymbol{\alpha}=(-1,-1)$, $\boldsymbol{\alpha}=(1,1)$]{\includegraphics[width=0.49\textwidth]{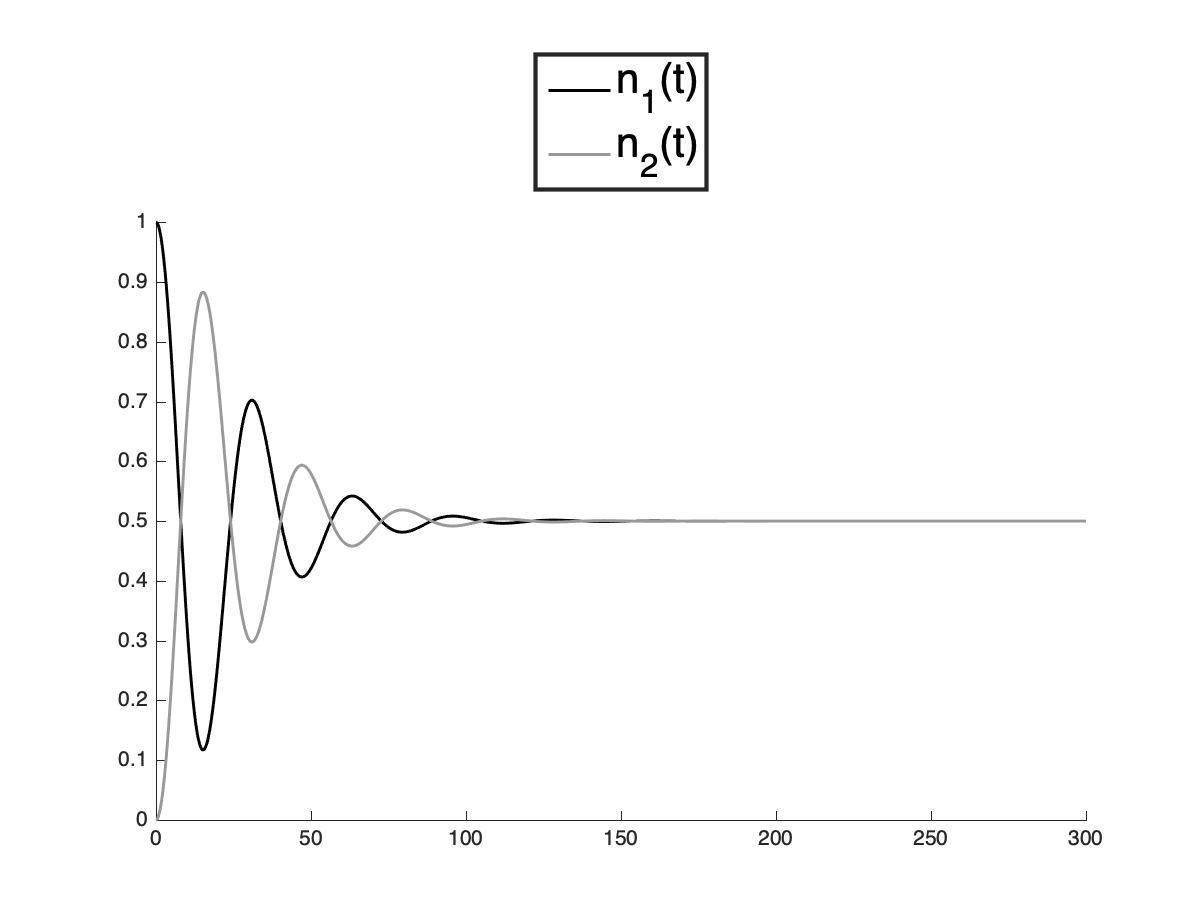}}
\subfigure[$\boldsymbol{\alpha}=(-1,1)$, $\boldsymbol{\alpha}=(1,-1)$]{\includegraphics[width=0.49\textwidth]{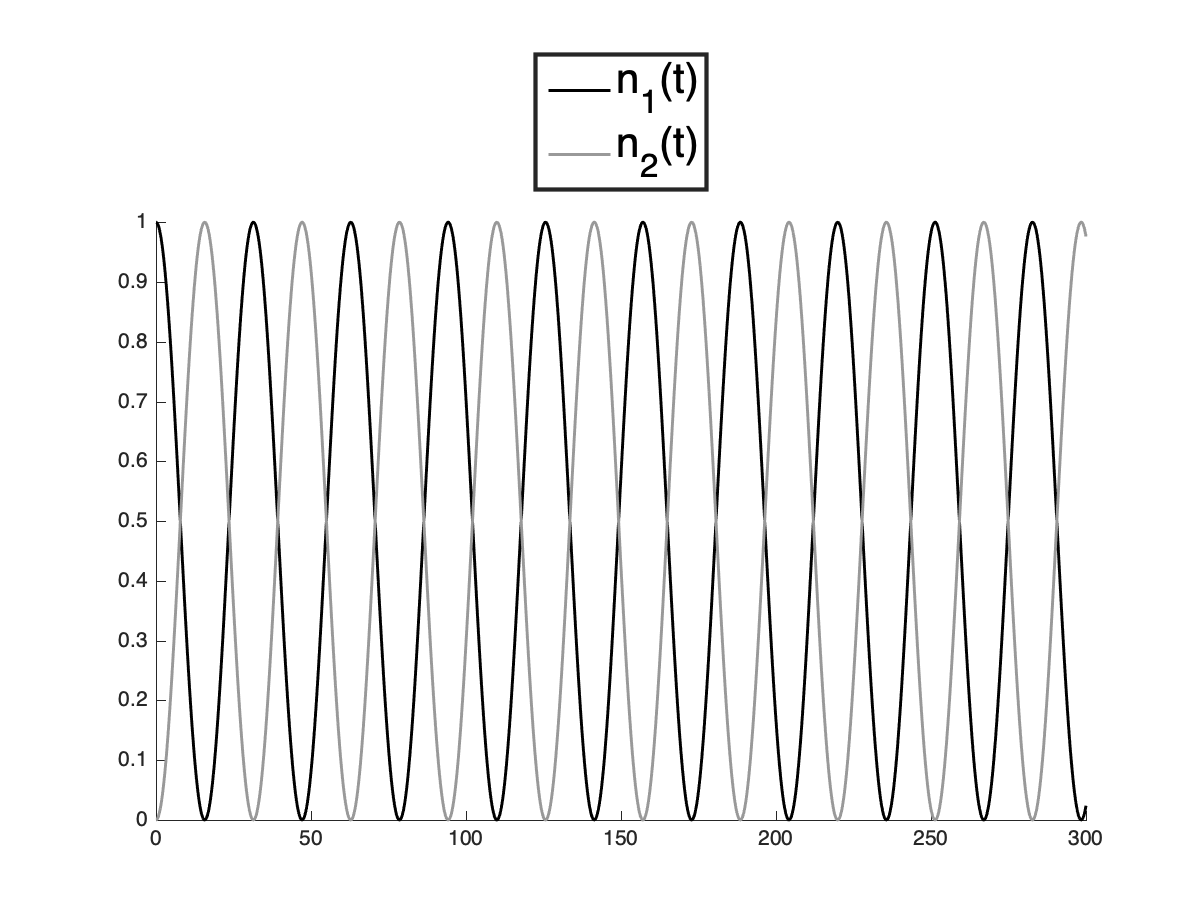}}\\
\caption{\label{fig_om1_eq_om2_lam_01}Plots of the solutions of equations \eqref{eq_rhocontinuous} with $\lambda=0.1$ and the initial values of the inertia parameters ${\omega_1}_0={\omega_2}_0=0.5$ according to various choices of $\alpha_1$ and $\alpha_2$.}
\end{figure}

\begin{table}
\caption{\label{tab_equil_a1}Asymptotic equilibrium state for $n_1(t)$; 
${\omega_1}_0=0.5$, ${\omega_2}_0=0.7$, $\lambda=0.1$, 
$\lambda=0.2$, and $\lambda=0.3$.}
\vspace{6pt}
\centering
\begin{tabular}{|c|c|c|c|c|}
\hline
$\alpha_1$ & $\alpha_2$ & $n_1^{eq}$ ($\lambda=0.1$) &  $n_1^{eq}$ ($\lambda=0.2$)  & $n_1^{eq}$ ($\lambda=0.3$) \\
\hline
0 & -1 & 0.1468 & 0.2771 & 0.3432\\
\hline
-1 & 0 & 0.1466 &0.2768 & 0.3424\\
\hline
0 & 1 & 0.8535 &0.7236 & 0.6581\\
\hline
1 & 0 & 0.8535 &0.7236 & 0.6581\\
\hline
-1 & -1 & 0.1466 &0.2769 & 0.3425\\
\hline
1 & 1 & 0.8535 & 0.7236 & 0.6581\\
\hline
-1 & 1 & 0.8536 &0.7236 & 0.6581\\
\hline
1 & -1 & 0.1468 &0.2771 & 0.3431\\
\hline
\end{tabular}
\end{table}

In this section, we introduce a suitable limit of the $(H,\rho)$-induced dynamics whereupon 
the definition of a new Hamiltonian operator, giving rise to a time evolution asymptotically approaching
an equilibrium state, is naturally suggested. Moreover, as it will be shown in the next section, 
a mathematical relation, linking the values of the parameters involved in the Hamiltonian with the value of the equilibrium state can be devised.

The rule \eqref{ruletau}  requires the choice of the length $\tau$ of the subinterval where we compute the 
variation of the state of the system which is needed to update the values of $\omega_1$ and $\omega_2$.
Different choices of $\tau$ imply different transient evolutions before the asymptotic equilibrium state
is reached; moreover, the choice of $\tau$ is somehow arbitrary.  
In order to have an evolution that is independent of $\tau$, once the initial condition $\varphi_{n_1,n_2}$ is chosen, we consider in a small interval $[t,t+\tau]$
the ratio
\begin{equation}
\frac{n_j(t+\tau)-n_j(t)}{\tau},\qquad j=1,2,
\end{equation}
\emph{i.e.}, the mean rate of change in the subinterval, and then take the limit for $\tau$ going to zero,
so obtaining the instantaneous rate of change of $n_j(t)$.

By using \eqref{dyneq2}, we have:
\begin{equation}
\dot {\widehat{n}}_1=-\dot {\widehat{n}}_2=\textrm{i}\lambda(a_1^\dagger a_2-a_2^\dagger a_1).
\end{equation}
Therefore, after computing the mean values $\dot n_i$ of $\dot {\widehat{n}}_i$ ($i=1,2$) on the initial state
$\varphi_{n_1,n_2}$, we may assume that at each instant of time $t$ the values of 
$\omega_1$ and $\omega_2$ are given by 
\begin{equation}
\label{rulecontinuous}
\begin{aligned}
&\omega_1={\omega_1}_0(1+\alpha_1 \dot n_1),\\
&\omega_2={\omega_2}_0(1+\alpha_2 \dot n_2),
\end{aligned}
\end{equation}
where ${\omega_1}_0$ and ${\omega_2}_0$ are the initial inertia parameters, whereas
$\alpha_1$ and $\alpha_2$ are some constants whose meaning is explained below.

Therefore, we assume the dynamics of our system to be ruled by the following set of nonlinear differential equations:
\begin{equation}
\label{eq_rhocontinuous}
\begin{aligned}
&\dot a_1=\textrm{i}\left(\lambda a_2-{\omega_1}_0(1+\textrm{i}\alpha_1 \lambda\langle \varphi_{n_1,n_2},
(a_1^\dagger a_2-a_2^\dagger a_1)\varphi_{n_1,n_2}\rangle) a_1\right),\\
&\dot a_2=\textrm{i}\left(\lambda a_1-{\omega_2}_0(1+\textrm{i}\alpha_2 \lambda\langle \varphi_{n_1,n_2},
(a_2^\dagger a_1-a_1^\dagger a_2)\varphi_{n_1,n_2}\rangle) a_2\right),
\end{aligned}
\end{equation}
to be solved with the initial condition 
\begin{equation}
a_1(0)=\left(
\begin{array}{cccc}
0 & 1 &0 & 0\\
0 & 0 & 0 & 0\\
0 & 0 & 0 & 1\\
0 & 0 & 0 & 0
\end{array}
\right),\qquad
a_2(0)=\left(
\begin{array}{cccc}
0 & 0 &1 & 0\\
0 & 0 & 0 & -1\\
0 & 0 & 0 & 0\\
0 & 0 & 0 & 0
\end{array}
\right),
\end{equation}
once the vector $\varphi_{n_1,n_2}$ has been fixed. Of course, we will be interested to the mean values of the occupation numbers on the initial condition given by $\varphi_{n_1,n_2}$.

The equations \eqref{eq_rhocontinuous} can be written formally as
\begin{equation}
\dot a_j=\textrm{i}[\widetilde{H},a_j], \qquad j=1,2,
\end{equation}
along with the \emph{generalized} Hamiltonian
\begin{equation}
\label{newHamiltonian}
\begin{aligned}
\widetilde{H}&={\omega_1}_0(1+\textrm{i}\alpha_1 \lambda\langle \varphi_{n_1,n_2},
(a_1^\dagger a_2-a_2^\dagger a_1)\varphi_{n_1,n_2}\rangle) a_1 a^\dagger_1\\
&+{\omega_2}_0(1-\textrm{i}\alpha_2 \lambda\langle \varphi_{n_1,n_2},
(a_1^\dagger a_2-a_2^\dagger a_1)\varphi_{n_1,n_2}\rangle) a_2 a^\dagger_2\\
&+\lambda(a_1 a^\dagger_2+a_2 a^\dagger_1),
\end{aligned}
\end{equation}
in whose definition we have included a term like
$$
\langle \varphi_{n_1,n_2},
(a_1^\dagger a_2-a_2^\dagger a_1)\varphi_{n_1,n_2}\rangle,
$$
which is proportional to the time derivative of the mean values of the occupation numbers on the assigned initial condition.
It is trivial to observe that
$$
\langle \varphi_{n_1,n_2},
(a_1(0)^\dagger a_2(0)-a_2(0)^\dagger a_1(0))\varphi_{n_1,n_2}\rangle=0,
$$
but, during the evolution ruled by \eqref{eq_rhocontinuous},
$$
\langle \varphi_{n_1,n_2},
(a_1(t)^\dagger a_2(t)-a_2(t)^\dagger a_1(t))\varphi_{n_1,n_2}\rangle\neq 0,
$$
approaching zero in a neighborhood of the equilibrium state.
This implies that the time evolution of $a_j(t)$ ($j=1,2$) can be obtained by integrating equations \eqref{eq_rhocontinuous} and not by computing 
$\exp(\textrm{i}\widetilde{H}t)a_j(0)\exp(-\textrm{i}\widetilde{H}t)$.

Notice also that, from equations \eqref{eq_rhocontinuous} it is easily derived that 
$n_1(t)+n_2(t)$ is a conserved quantity.

The constants $\alpha_1$ and $\alpha_2$ are related to the effects we want to model. We can choose their values in order to take into account into the Hamiltonian only one between 
$\dot n_1$ and $\dot n_2$ ($\alpha_1\alpha_2=0$, $\alpha_1+\alpha_2\neq 0$), or both
($\alpha_1\alpha_2\neq 0$); 
if $\alpha_1\alpha_2>0$  the coefficients in the free part of $H$ in \eqref{newHamiltonian} change in the opposite way, since $\dot n_1(t)=-\dot n_2(t)$; on the contrary, they change in the same way if 
($\alpha_1\alpha_2<0$). Of course, if $\alpha_1=\alpha_2=0$, we reduce to the standard 
case considered in the previous section.

To keep the situation simple, let us take, without loss of generality, $\alpha_1,\alpha_2 \in \{-1,0,1\}$. In fact, the numerical solutions show that the values of the asymptotic states are not changed by choosing
non zero values of $\alpha_1$ and $\alpha_2$ different from $\{-1,1\}$.

The numerical integration of \eqref{eq_rhocontinuous} for various choices of the values of 
${\omega_1}_0$,  ${\omega_2}_0$ and $\lambda$, as well as the parameters 
$\alpha_1,\alpha_2$, show that the system, after a transient dynamics, excluding very few
particular situations whose aspects will be discussed in the following, always approaches  an asymptotic equilibrium state. Of course, for values of $t$ large enough, when the equilibrium state is reached, the contribution $\langle \varphi_{n_1,n_2},
(a_1(t)^\dagger a_2(t)-a_2(t)^\dagger a_1(t))\varphi_{n_1,n_2}\rangle$ is vanishing.
 
In the simulations shown in this paper, we used the initial condition $\varphi_{1,0}$, \emph{i.e.}, 
the initial mean values $n_1=1$, $n_2=0$. This is not limiting because we consider all the possible cases: ${\omega_1}_0<{\omega_2}_0$, ${\omega_2}_0<{\omega_1}_0$, and ${\omega_1}_0={\omega_2}_0$.

Figure~\ref{fig_om1_lt_om2_lam_01} clearly shows that our model more or less quickly reaches its equilibrium state 
in the cases where $\alpha_1\alpha_2=0,1$. As already remarked, the asymptotic equilibrium 
$n_1^{eq}$ does not change when the non zero values of $\alpha_1$ and $\alpha_2$ do not belong to 
the set $\{-1,1\}$: what changes is only the transient evolution. As shown in Table~\ref{tab_equil_a1}, 
the asymptotic equilibrium $n_1^{eq}$ for the parameters $(\alpha_1,\alpha_2)$ is the complement to 1 
of the one for the parameters $(-\alpha_1,-\alpha_2)$, \emph{i.e.}, it depends on
 $\hbox{sign}(\alpha_1+\alpha_2)$ if the latter is different from zero. 

In the cases where $\alpha_1\alpha_2=-1$ (this means that the coefficients of the free part of $H$ both increase or decrease during the evolution), the equilibrium state is reached after a longer time, as it is evident in  Figure~\ref{fig_om1_lt_om2_lam_01_bis}, where a larger time range needs to be considered in order to go beyond the transient period.

Choosing ${\omega_1}_0=0.7$, ${\omega_2}_0=0.5$, the time evolutions we get for 
$\alpha_1\alpha_2\neq -1$, as one would expect, are those depicted in Figure~\ref{fig_om1_lt_om2_lam_01}, provided that we change the signs of $\alpha_1$ and $\alpha_2$; for instance, if 
${\omega_1}_0>{\omega_2}_0$, the evolution in the case 
$\boldsymbol\alpha=(0,-1)$ is that obtained for ${\omega_1}_0<{\omega_2}_0$, in the case 
$\boldsymbol\alpha=(0,1)$, and so on.

On the contrary, if $\alpha_1\alpha_2=-1$, we get the same time evolution when we exchange the initial values of ${\omega_1}_0$ and
${\omega_2}_0$.

The same qualitative considerations apply if we change the value of $\lambda$. What it can be observed is that higher values of $\lambda$ 
determine a shorter time to reach the equilibrium state 
(see Figures~\ref{fig_om1_lt_om2_lam_03} and 
\ref{fig_om1_lt_om2_lam_03_bis}); this is reasonable, since the value of $\lambda$ determines the
strength of the interaction and the term 
$\langle \varphi_{n_1,n_2},
(a_1^\dagger a_2-a_2^\dagger a_1)\varphi_{n_1,n_2}\rangle$ in the dynamical equations \eqref{eq_rhocontinuous} is multiplied by $\lambda$;
moreover, also the value of the asymptotic equilibrium changes with $\lambda$ (see Table~\ref{tab_equil_a1}).  

The case where ${\omega_1}_0={\omega_2}_0$ is much simpler. In fact,
the time evolution for $\boldsymbol\alpha=(\alpha_1,\alpha_2)$  
is the same as that obtained for $\boldsymbol\alpha=(-\alpha_1,-\alpha_2)$. The cases where $\alpha_1\alpha_2=-1$ produce the particular situations we mentioned before in which the time evolution remains periodic ($n_1(t)$ and $n_2(t)$ oscillate in the whole interval $[0,1]$), and no asymptotic equilibrium is reached by the system. This is not surprising, since we start with equal inertia parameters and their variation during the evolution is always in the same direction. We also observe that the asymptotic equilibrium state is always $n_1^{eq}=0.5$; when the solution is periodic, 0.5 is the integral mean of the solution itself.

A final comment is in order. When the evolution of the system is very close to the equilibrium state, the Hamiltonian \eqref{newHamiltonian}, since $\dot n_1=\dot n_2\approx 0$, tends to \eqref{hamiltonian}. The latter, due to its structure, is such that $n_1(t)$ and $n_2(t)$ should oscillate periodically in opposition of phase; nevertheless, $n_1(t)$ and $n_2(t)$ do not. Therefore, we can say that a synchronization of the oscillations of $n_1(t)$ and $n_2(t)$ has emerged, whereupon, since  $n_1(t)+n_2(t)$ is a conserved quantity, an equilibrium state arises.

\section{Asymptotic steady states and parameters}
\label{sect4}

\begin{figure}[!h]
\centering
\subfigure[$\boldsymbol\alpha=(0,-1)$]{\includegraphics[width=0.49\textwidth]{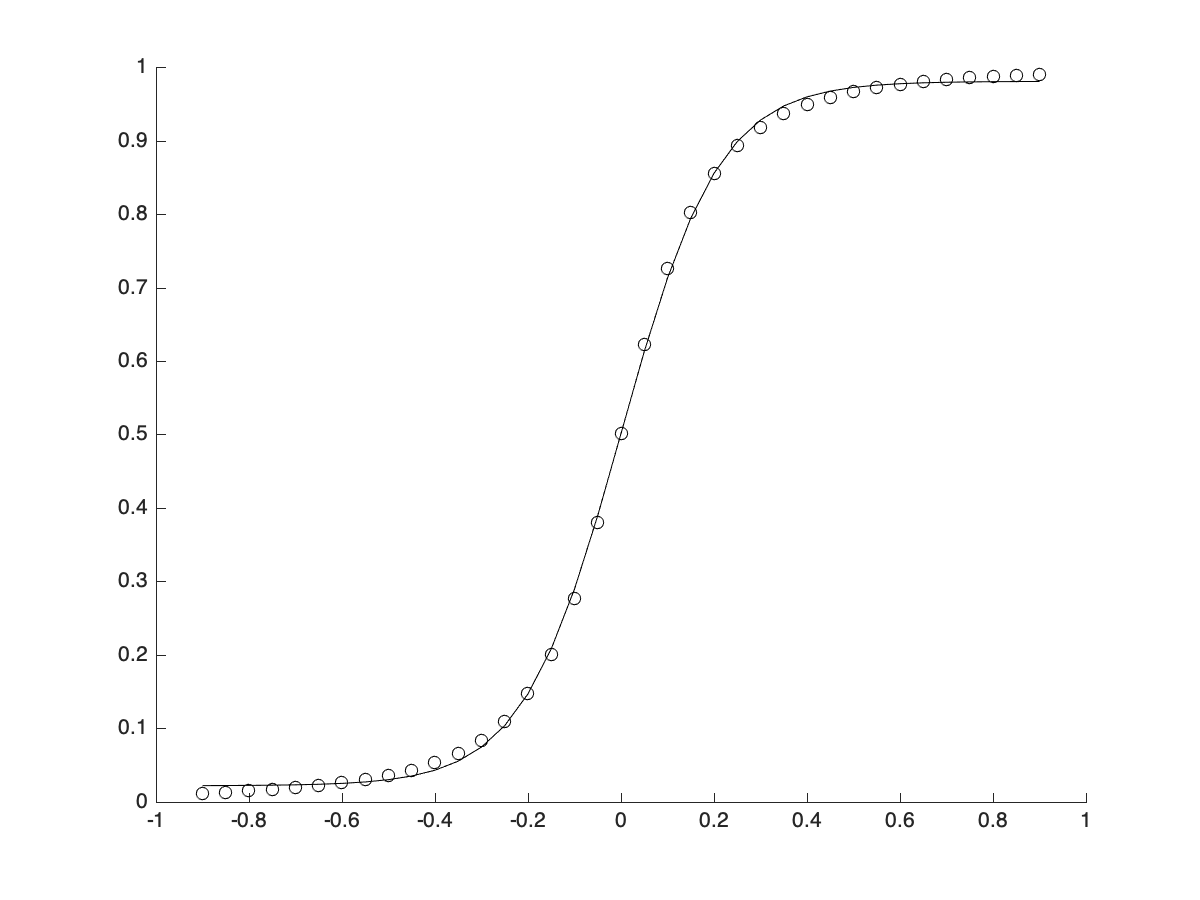}}
\subfigure[$\boldsymbol\alpha=(0,1)$]{\includegraphics[width=0.49\textwidth]{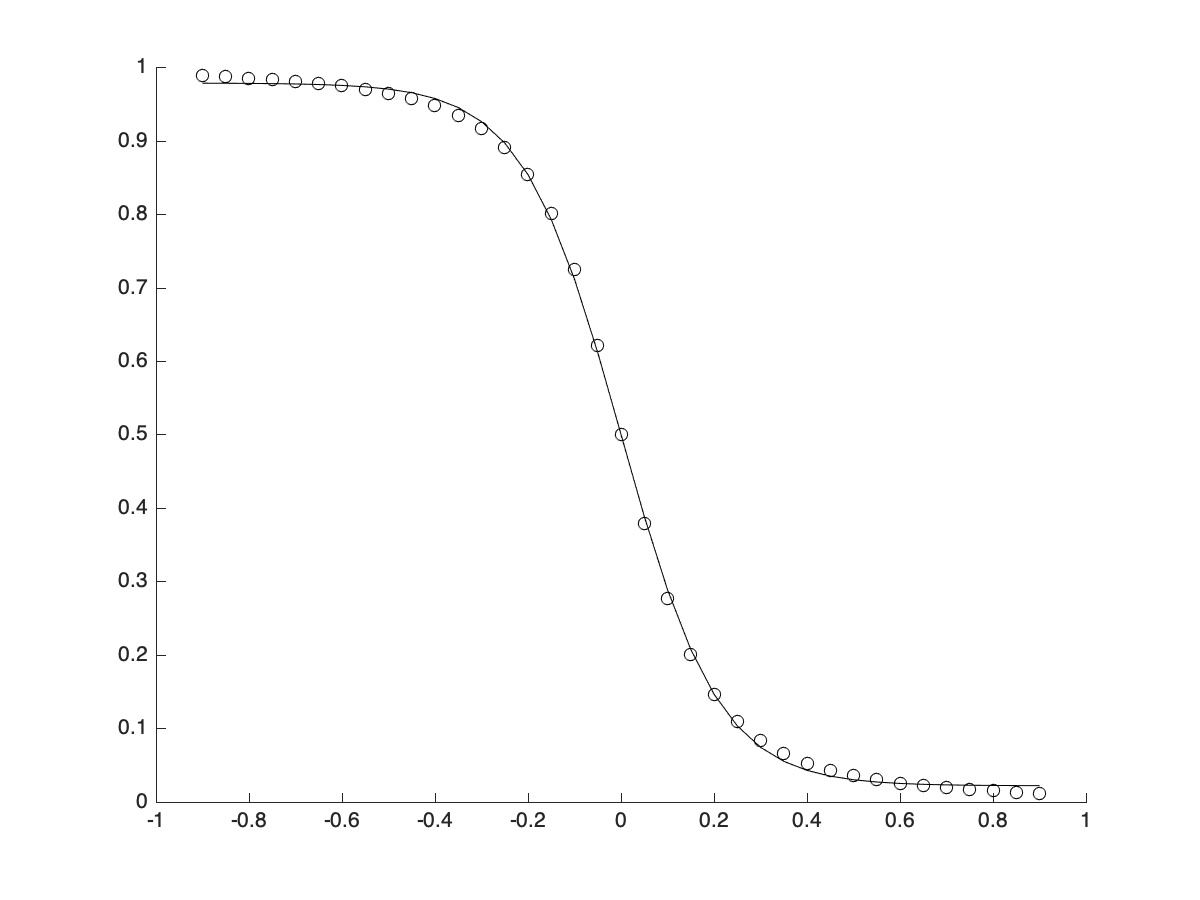}}
\caption{\label{fig_fitting}Equilibrium values of $n_1$ vs. ${\omega_1}_0-{\omega_2}_0$ for $\lambda=0.1$, fitted by means of function \eqref{fitting}.} 
\end{figure}

\begin{figure}
\centering
\subfigure[$\boldsymbol\alpha=(-1,1)$]{\includegraphics[width=0.49\textwidth]{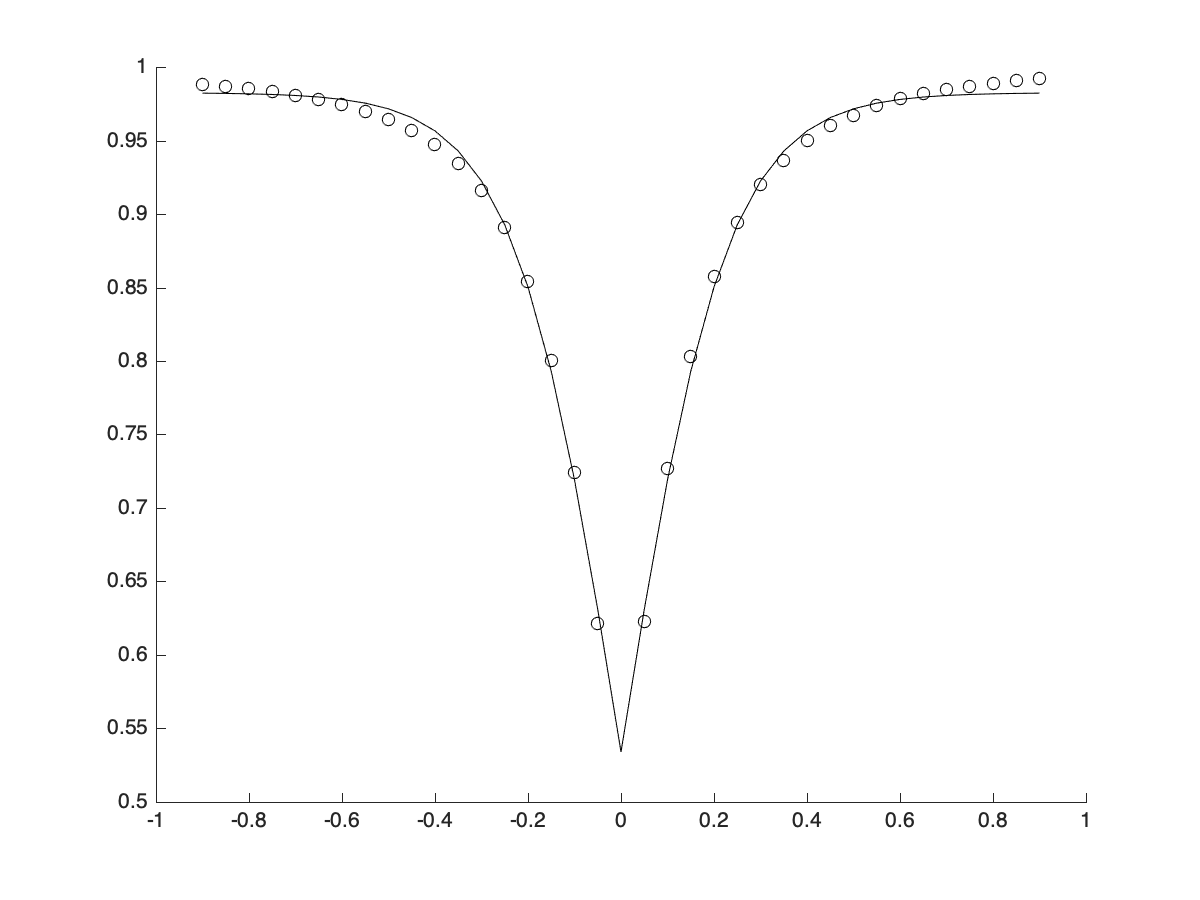}}
\subfigure[$\boldsymbol\alpha=(1,-1)$]{\includegraphics[width=0.49\textwidth]{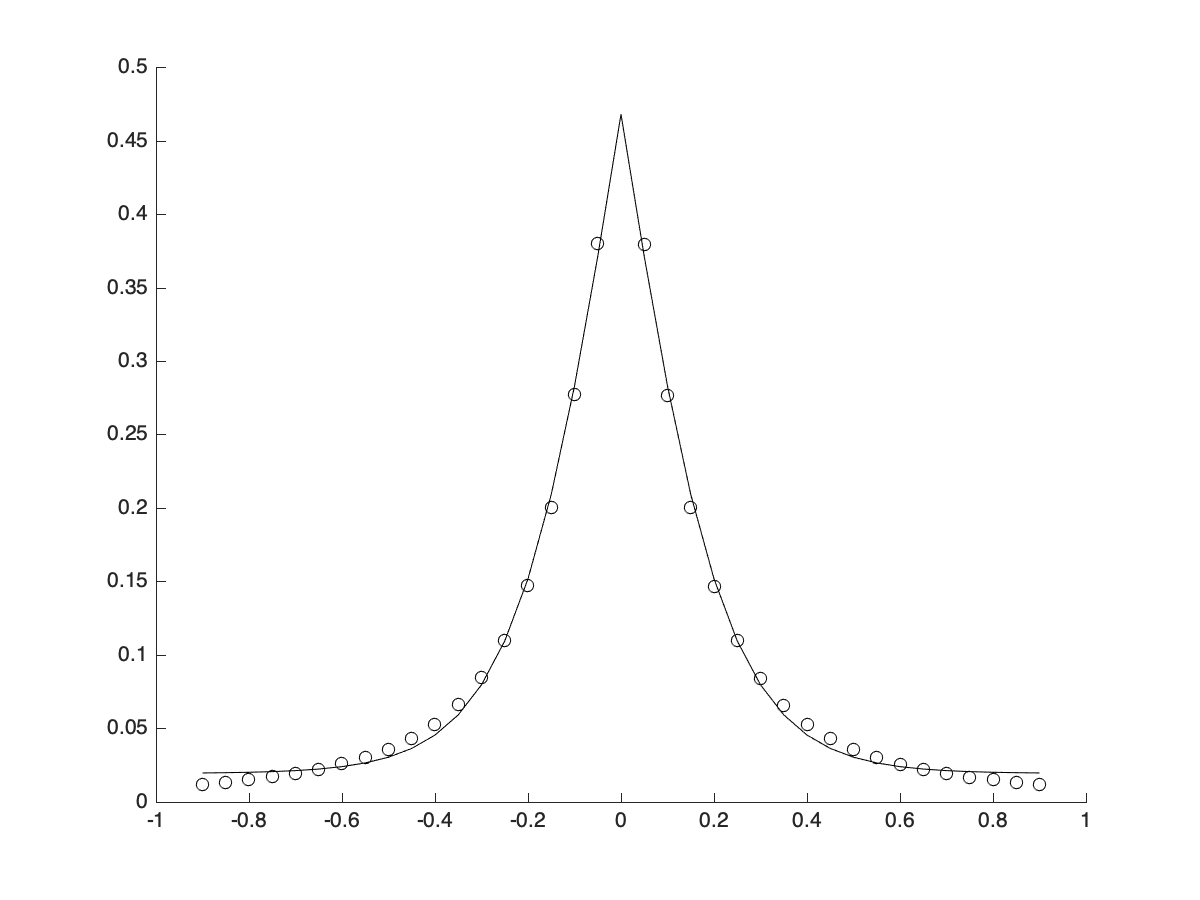}}
\caption{\label{fig_fittingbis}Equilibrium values of $n_1$ vs. ${\omega_1}_0-{\omega_2}_0$ for $\lambda=0.1$, fitted by means of function \eqref{relation2}.} 
\end{figure}

In this section, we show that, independently of the choice of $(\alpha_1,\alpha_2)$, a mathematical relation can be found giving the value of the asymptotic equilibrium state for $n_1(t)$ (which of course determines the asymptotic equilibrium value of $n_2(t)$ because of the constraint $n_1(t)+n_2(t)=\hbox{constant}$)
as a function of ${\omega_1}_0$, ${\omega_2}_0$, $\lambda$, $\alpha_1$ and $\alpha_2$.
\begin{table}
\caption{\label{tab_regression}Parameters entering the fitting \eqref{fitting} as a function of $\lambda$.}
\vspace{6pt}
\centering
\begin{tabular}{|c|c|c|c||c|c|c|c|}
\hline
$\lambda$ & $a$ & $b$ & $c$ & $\lambda$ & $a$ & $b$ & $c$\\
\hline
0.05 & 0.5012 & 0.4903 & 8.9566 & 0.55  &0.5021 & 0.4231 & 1.0733  \\
\hline
0.1 & 0.5013 & 0.4793 & 4.7420 & 0.6  & 0.5020 & 0.4229 & 0.9832\\
\hline
0.15 & 0.5014 & 0.4688 & 3.3296 & 0.65  & 0.5022 & 0.4222 & 0.9092\\
\hline
0.2 & 0.5015 & 0.4595 & 2.6037 & 0.7  & 0.5024 & 0.4167 & 0.8582\\
\hline
0.25 & 0.5015 & 0.4515 & 2.1494 & 0.75  & 0.5024 & 0.4147 & 0.8062\\  
\hline
0.3  & 0.5016 & 0.4457 & 1.8292 & 0.8  & 0.5018 & 0.4154 & 0.7516 \\
\hline
0.35 & 0.5018 & 0.4395 & 1.6031 & 0.85  & 0.5021 & 0.4140 & 0.7116 \\
\hline
0.4  & 0.5017 & 0.4354 & 1.4215 & 0.9  & 0.5022 & 0.4047 & 0.6904\\
\hline
0.45  & 0.5017 & 0.4310 & 1.2816 & 0.95  & 0.5023 & 0.4053 & 0.6534\\
\hline
0.5  & 0.5019 & 0.4251 & 1.1761 &  1.0  & 0.5024 & 0.4168 & 0.6013 \\
\hline
\end{tabular}
\vspace*{-4pt}
\end{table}

Let us fix ${\omega_1}_0=1$ and compute the value $n_1^{eq}$ varying 
${\omega_2}_0$ between 0.1 and 1.9, for $\lambda=0.1$. We start considering the case where $\boldsymbol\alpha=(0,-1)$.

The obtained data suggest to look for a nonlinear fit with the function
\begin{equation}
\label{fitting}
f(x)=a+b\, \hbox{tanh}(cx),
\end{equation}
where $x={\omega_1}_0-{\omega_2}_0$, and $a$, $b$ and $c$ are constant. The data are well fitted by using \eqref{fitting} along with the parameters
\begin{equation}
a = 0.5013, \quad b=0.4793, \quad c=4.7420.
\end{equation}
If we consider the case $\boldsymbol\alpha=(0,1)$, function \eqref{fitting} also well reproduces the equilibrium values $n_1^{eq}$ by using the parameters 
\begin{equation}
a=0.5001,\quad b =- 0.4782, \quad c =4.7458.
\end{equation}
Both situations are depicted in Figure~\ref{fig_fitting}.

Moreover, if we look at the values of $a$, $b$ and $c$ obtained in the case $\boldsymbol\alpha=(0,-1)$ in correspondence to the different values of $\lambda$ (reported in Table~\ref{tab_regression}), we observe that the parameter $a$ do not seem to depend on $\lambda$, whereas $b$ depends weakly upon 
$\lambda$, and $c$ is strongly affected by the value of $\lambda$. Since the equilibrium value in the case $\boldsymbol\alpha =(0,1)$ corresponds to the complement to 1
of the equilibrium value in the case $\boldsymbol\alpha =(0,-1)$, we should have exactly $a=1/2$. 
Moreover, neglecting the unavoidable numerical errors in the nonlinear fitting, it is reasonable to 
assume $b = 1/2$ in \eqref{fitting}.

Along with these assumptions, looking for a fit relating $c$ to $\lambda$, we obtain the following relation:
\[
c=\frac{0.5107}{\lambda}\approx \frac{1}{2\lambda}.
\]

Therefore, we assume in the case $\boldsymbol\alpha=(0,-1)$ that the law
\begin{equation}
\label{regression1}
n_1^{eq}=\frac{1}{2}\left(1+\hbox{tanh}\left(\frac{{\omega_1}_0-{\omega_2}_0}{2\lambda}\right)\right)
\end{equation}
provides the relation between the constant parameters entering the Hamiltonian 
\eqref{newHamiltonian} and $n_1^{eq}$. By setting
\[
\mu=\frac{{\omega_1}_0-{\omega_2}_0}{2\lambda}
\]
we also notice that the quantity $\delta$ in \eqref{period}$_1$ (related to the periodicity of the evolution
when $\boldsymbol\alpha=(0,0)$) can be written as
\[
\delta=2\lambda\sqrt{\mu^2+1}.
\]

It can be verified that the relation \eqref{regression1} holds also when 
$\boldsymbol\alpha=(-1,0)$ and $\boldsymbol\alpha=(-1,-1)$. On the contrary, in the cases
$\boldsymbol\alpha=(0,1)$, $\boldsymbol\alpha=(1,0)$ and $\boldsymbol\alpha=(1,1)$ the relation giving the value of the asymptotic equilibrium state reads
\begin{equation}
\label{regression2}
n_1^{eq}=\frac{1}{2}\left(1-\hbox{tanh}\left(\frac{{\omega_1}_0-{\omega_2}_0}{2\lambda}\right)\right).
\end{equation}

Therefore, in all the cases where $\alpha_1\alpha_2\neq -1$, the relation giving the value of $n_1^{eq}$ in terms of the initial parameters entering the Hamiltonian can be written in the unified way
\begin{equation}
\label{regression}
n_1^{eq}=\frac{1}{2}\left(1-\hbox{sign}(\alpha_1+\alpha_2)\hbox{tanh}\left(\frac{{\omega_1}_0-{\omega_2}_0}{2\lambda}\right)\right).
\end{equation}

The situation is a little bit different when $\alpha_1\alpha_2=-1$, where the formula 
\begin{equation}
\label{relation2}
{n_1}^{eq}=\frac{1}{2} \left( 1-\hbox{sign}(\alpha_1)\hbox{sign}({\omega_1}_0-{\omega_2}_0)\hbox{tanh}\left(\frac{{\omega_1}_0-{\omega_2}_0}{2\lambda}\right)\right),
\end{equation}
giving the equilibrium state for $n_1(t)$ (see Figure~\ref{fig_fittingbis}), can be derived.

As already pointed out, when $\alpha_1\alpha_2=-1$, and ${\omega_1}_0={\omega_2}_0$, the evolution remains periodic, with $n_1(t)$ and $n_2(t)$ oscillating between 0 and 1 without approaching any equilibrium state. Nevertheless, formula \eqref{relation2} predicts an equilibrium value equal to 
$\frac{1}{2}$: the formula can be considered coherent also in this case since the mean value of $n_1(t)$ (oscillating between 0 and 1) is exactly $\frac{1}{2}$.
At the present, we do not have a rigorous argument in order to justify mathematically the relation between the parameters of the
model and the value of the equilibrium; further investigations are planned, also considering more complicated systems and Hamiltonians.

\section{Conclusions and perspectives}
\label{sect5}

In some previous papers, we have discussed how the approach of the $(H,\rho)$-induced dynamics can be quite efficient to produce dynamical behaviors admitting some asymptotic non trivial limits, without the need of introducing any infinitely extended reservoir coupled to the system, or any explicitly time dependent Hamiltonian. By considering a suitable limit in the $(H,\rho)$-induced approach, we continuously act on the parameters embodied in the Hamiltonian, and define a generalized Hamiltonian operator giving rise to a time evolution approaching specific equilibria. This generalized operator is time independent and self-adjoint, but in the parameters entering its free part a dependence on the time derivative of the mean values on a given eigenstate of the number operators is embedded.
We remark that the sequence of the Hamiltonian operators ruling the dynamics according to the $(H,\rho)$ approach can be viewed as a discrete version of the generalized Hamiltonian introduced in this paper.
Within this generalized framework we are able to describe inside the usual Heisenberg view asymptotic behaviors, even in the case of a two-mode fermionic system living in a finite-dimensional Hilbert space. Moreover, we provide a relation linking the value of the equilibria to the set of the parameters of the model. This allows us to tune the parameters of the model in order to obtain a given asymptotic limit.

Work is in progress with the aim of applying this generalized approach to concrete situations well described by operatorial models with more than two fermionic modes.

\section*{Acknowledgments}
The authors acknowledge partial support from G.N.F.M. of ``Istituto Nazionale di Alta Matematica''. The authors
are grateful to the unknown referees for their valuable comments.

\end{document}